\documentclass[12pt]{article}
\usepackage{amsmath}
\usepackage{amssymb}
\usepackage{myart}

\normalbaselines
\input tcilatex
\begin{document}

\title{Geometrical dynamics: spin as a result of rotation with superluminal
speed.}
\author{Yuri A.Rylov}
\date{Institute for Problems in Mechanics, Russian Academy of Sciences,\\
101-1, Vernadskii Ave., Moscow, 119526, Russia.\\
e-mail: rylov@ipmnet.ru\\
Web site: {$http://rsfq1.physics.sunysb.edu/\symbol{126}rylov/yrylov.htm$}\\
or mirror Web site: {$http://gasdyn-ipm.ipmnet.ru/\symbol{126}%
rylov/yrylov.htm$}}
\maketitle

\begin{abstract}
Dynamics is considered as a corollary of the space-time geometry. Evolution
of a particle in the space-time is described as a chain of connected
equivalent geometrical objects. Space-time geometry is determined uniquely
by the world function $\sigma $. Proper modification of the Minkowskian
world function for large space-time interval leads to wobbling of the chain,
consisted of timelike straight segments. Statistical description of the
stochastic world chain coincides with the quantum description by means of
the Schr\"{o}dinger equation. Proper modification of the Minkowskian world
function for small space-time interval may lead to appearance of a world
chain, having a shape of a helix with timelike axis. Links of the chain are
spacelike straight segments. Such a world chain describes a spatial
evolution of a particle. In other words, the helical world chain describes
the particle rotation with superluminal velocity. The helical world chain
associated with the classical Dirac particle, whose world line is a helix.
Length of world chain link cannot be arbitrary. It is determined by the
space-time geometry and, in particular, by the elementary length. There
exists some discrimination mechanism, which can discriminate some world
chains.
\end{abstract}

\section{Introduction}

Geometrical dynamics is a dynamics of elementary particles, generated by the
space-time geometry. In the space-time of Minkowski the geometrical dynamics
coincides with the conventional classical dynamics, and the geometrical
dynamics may be considered to be a generalization of classical dynamics onto
more general space-time geometries. However, the geometric dynamics has more
fundamental basis, and it may be defined in multivariant space-time
geometries, where one cannot introduce the conventional classical dynamics.
The fact is that, the classical dynamics has been introduced for the
space-time geometry with unlimited divisibility, whereas the real space-time
has a limited divisibility. The limited divisibility of the space-time is of
no importance for dynamics of macroscopic bodies. However, when the size of
moving bodies is of the order of the size of heterogeneity, one may not
neglect the limited divisibility of the space-time geometry.

The geometric dynamics is developed in the framework of the program of the
further physics geometrization, declared in \cite{R2007a}. The special
relativity and the general relativity are steps in the development of this
program. Necessity of the further development appeared in the thirtieth of
the twentieth century, when diffraction of electrons has been discovered.
The motion of electrons, passing through the narrow slit, is multivariant.
As far as the free electron motion depends only on the properties of the
space-time, one needed to change the space-time geometry, making it to be
multivariant. In multivariant geometry there are many vectors $\mathbf{Q}_{0}%
\mathbf{Q}_{1}$, $\mathbf{Q}_{0}\mathbf{Q}_{1}^{\prime }$,...at the point $%
Q_{0}$, which are equal to the vector $\mathbf{P}_{0}\mathbf{P}_{1}$, given
at the point $P_{0}$, but they are not equal between themselves, in general.
Such a space-time geometry was not known in the beginning of the twentieth
century. It is impossible in the framework of the Riemannian geometry. As a
result the multivariance was prescribed to dynamics. To take into account
multivariance, dynamic variables were replaced by matrices and operators.
One obtains the quantum dynamics, which differs from the classical dynamics
in its principles. But the space-time conception remains to be Newtonian
(nonrelativistic). Multivariant space-time geometry appeared only in the end
of the twentieth century \cite{R90,R2001}. The further geometrization of
physics became to be possible.

It should note that there were numerous attempts of further geometrization
of physics. They were based on the Riemannian space-time geometry.
Unfortunately, the true space-time geometry of microcosm does not belong to
the class of Riemannian geometries, and approximation of real space-time
geometry by a Riemannian geometry cannot be completely successful. In
particular, the Riemannian geometry cannot describe such a property of real
space-time geometry as multivariance. The multivariance of the space-time
geometry was replaced by the multivariance of dynamics (quantum theory).

Understanding of nature of elementary particles is the aim of the further
geometr\TEXTsymbol{\backslash}-ization of physics. This aim distinguishes
from the aim of the conventional theory of elementary particles. Let us
explain the difference of aims in the example of history the chemical
elements investigation. Investigation of chemical elements reminds to some
extent investigation of elementary particles. Chemical elements are
investigated from two sides. Chemists systematized chemical elements,
investigating their phenomenological properties. The results of these
investigations were formulated in the form of the periodical system of
chemical elements in 1870. Formulating this system, D.I.Mendeleev conceived
nothing about the atom construction. Nevertheless the periodical system
appears to be very useful from the practical viewpoint. Physicists did not
aim to explain the periodical system of chemical elements, they tried to
understand simply the atom structure and the discrete character of atomic
spectra. After construction of the atomic theory it became clear, that the
periodical system of chemical elements can be obtained and explained on the
basis of the atomic theory. As a result the "physical" approach to
investigation of chemical elements appeared to be more fundamental, deep and
promising, than the "chemical" one. On the other hand, the way of the
"physical" approach to explanation of the periodical system is very long and
difficult. Explanation of the periodical system was hardly possible at the
"physical" approach, i.e. without the intermediate aim (construction of
atomic structure).

Thus, using geometrization of physics, we try to approach only intermediate
aim: explanation of multivariance of particle motion (quantum motion) and
capacity of discrimination of particle masses. Discrete character of masses
of elementary particles can be understood, only if we understand the reason
of the elementary particle discrimination. Contemporary approach to the
elementary particle theory is the "chemical" (phenomenological) approach. It
is useful from the practical viewpoint. However, it admits hardly to
understand nature of elementary particles, because the nature of the
discrimination mechanism, leading to discrete characteristics of elementary
particle, remains outside the consideration.

The most general geometry is a physical geometry, which is called also the
tubular geometry (T-geometry) \cite{R90,R2001,R2005}, because straights in
T-geometry are hallow tubes, in general. The T-geometry is determined
completely by its world function $\sigma \left( P,Q\right) =\frac{1}{2}\rho
^{2}\left( P,Q\right) $, where $\rho \left( P,Q\right) $ is interval between
the points $P$ and $Q$ in space-time, described by the T-geometry. All
concepts of T-geometry are expressed in terms of the world function $\sigma $%
. Dynamics of particles (geometric dynamics) is also described in terms of
the world function. The elementary particle is considered as an elementary
geometrical object (EGO) in the space-time. The elementary geometrical
object $\mathcal{O}$ is described by its skeleton $\mathcal{P}^{n}=\left\{
P_{0},P_{1},...P_{n}\right\} $ and its envelope $\mathcal{E}$. The envelope $%
\mathcal{E}$ is defined as a set of zeros of the envelope function $f_{%
\mathcal{P}^{n}}$%
\begin{equation}
\mathcal{O}=\left\{ R|f_{\mathcal{P}^{n}}\left( R\right) =0\right\}
\label{c1.1}
\end{equation}%
The envelope function $f_{\mathcal{P}^{n}}$ is a real function of arguments $%
w=\left\{ w_{1},w_{2},...w_{s}\right\} $. Any argument $w_{k}$ $k=1,2,...s$
is a world function $w_{k}=\sigma \left( L_{k},S_{k}\right) $, $%
L_{k},S_{k}\in \left\{ R,\mathcal{P}^{n}\right\} $. It is supposed that EGO
with skeleton $\mathcal{P}^{n}=\left\{ P_{0},P_{1},...P_{n}\right\} $ is
placed at the point $P_{0}$.

In T-geometry the vector $\overrightarrow{P_{0}P_{1}}\equiv \mathbf{P}_{0}%
\mathbf{P}_{1}$ is an ordered set of two points $\left\{ P_{0},P_{1}\right\} 
$. The length $\left\vert \mathbf{P}_{0}\mathbf{P}_{1}\right\vert $ of the
vector $\mathbf{P}_{0}\mathbf{P}_{1}$ is defined via the world function by
means of the relation%
\begin{equation}
\left\vert \mathbf{P}_{0}\mathbf{P}_{1}\right\vert ^{2}=2\sigma \left(
P_{0},P_{1}\right)  \label{c1.2}
\end{equation}

The scalar product $\left( \mathbf{P}_{0}\mathbf{P}_{1}.\mathbf{Q}_{0}%
\mathbf{Q}_{1}\right) $ of two vectors $\mathbf{P}_{0}\mathbf{P}_{1}$ and $%
\mathbf{Q}_{0}\mathbf{Q}_{1}$ is defined by the relation%
\begin{equation}
\left( \mathbf{P}_{0}\mathbf{P}_{1}.\mathbf{Q}_{0}\mathbf{Q}_{1}\right)
=\sigma \left( P_{0},Q_{1}\right) +\sigma \left( P_{1},Q_{0}\right) -\sigma
\left( P_{0},Q_{0}\right) -\sigma \left( P_{1},Q_{1}\right)  \label{c1.3}
\end{equation}%
Equivalence $\mathbf{P}_{0}\mathbf{P}_{1}$eqv$\mathbf{Q}_{0}\mathbf{Q}_{1}$
of two vectors $\mathbf{P}_{0}\mathbf{P}_{1}$ and $\mathbf{Q}_{0}\mathbf{Q}%
_{1}$ is defined as follows. Two vectors $\mathbf{P}_{0}\mathbf{P}_{1}$ and $%
\mathbf{Q}_{0}\mathbf{Q}_{1}$ are equivalent (equal), if%
\begin{equation}
\mathbf{P}_{0}\mathbf{P}_{1}\mathrm{eqv}\mathbf{Q}_{0}\mathbf{Q}_{1}:\quad
\left( \left( \mathbf{P}_{0}\mathbf{P}_{1}.\mathbf{Q}_{0}\mathbf{Q}%
_{1}\right) =\left\vert \mathbf{P}_{0}\mathbf{P}_{1}\right\vert \cdot
\left\vert \mathbf{Q}_{0}\mathbf{Q}_{1}\right\vert \right) \wedge \left(
\left\vert \mathbf{P}_{0}\mathbf{P}_{1}\right\vert =\left\vert \mathbf{Q}_{0}%
\mathbf{Q}_{1}\right\vert \right)  \label{c1.4}
\end{equation}%
In the developed form we have%
\begin{eqnarray*}
\sigma \left( P_{0},Q_{1}\right) +\sigma \left( P_{1},Q_{0}\right) -\sigma
\left( P_{0},Q_{0}\right) -\sigma \left( P_{1},Q_{1}\right) &=&2\sigma
\left( P_{0},P_{1}\right) \\
\sigma \left( P_{0},P_{1}\right) &=&\sigma \left( Q_{0},Q_{1}\right)
\end{eqnarray*}

Skeletons $\mathcal{P}^{n}=\left\{ P_{0},P_{1},...P_{n}\right\} $ and $%
\mathcal{Q}^{n}=\left\{ Q_{0},Q_{1},...Q_{n}\right\} $ are equivalent ($%
\mathcal{P}^{n}$eqv$\mathcal{Q}^{n})$, if corresponding vectors of both
skeletons are equivalent%
\begin{equation}
\mathcal{P}^{n}\mathrm{eqv}\mathcal{Q}^{n}:\quad \mathbf{P}_{i}\mathbf{P}_{k}%
\mathrm{eqv}\mathbf{Q}_{i}\mathbf{Q}_{k},\qquad i,k=0,1,...n  \label{c1.5}
\end{equation}

The skeleton $\mathcal{P}^{n}=\left\{ P_{0},P_{1},...P_{n}\right\} $ of EGO
at the point $P_{0}$ may exist as a skeleton of a physical body, if it may
exist at any point $Q_{0}\in \Omega $ of the space-time $\Omega $. It means
that there is a solution for system of equations 
\begin{equation}
\mathbf{P}_{i}\mathbf{P}_{k}\mathrm{eqv}\mathbf{Q}_{i}\mathbf{Q}_{k},\qquad
i,k=0,1,...n  \label{c1.6}
\end{equation}%
for any point $Q_{0}\in \Omega $. Further for brevity we take, that an
existence of a skeleton means an existence of corresponding geometrical
object.

In the space-time of Minkowski the problem of the skeleton existence is
rather simple, because at given $\mathcal{P}^{n}$ and $Q_{0}$ the system (%
\ref{c1.6}) of $n\left( n+1\right) $ algebraic equations has a unique
solution, although the number of equations may distinguish from the number
of variables to be determined. Indeed, in the four-dimensional space-time
the number of coordinates of $n$ points $Q_{1},Q_{2},...Q_{n}$ is equal to $%
4n$ (the point $Q_{0}$ is supposed to be given). If $n>3,$ the number $%
n\left( n+1\right) $ of equations is larger than the number ($4n$) of
variables. In the case of an arbitrary space-time geometry (arbitrary world
function $\sigma $) existence of solution of the system (\ref{c1.6}) is
problematic, and the question of existence of the skeleton as a skeleton of
a physical body is an essential problem. On the contrary, if $n<3$, the
number of coordinates to be determined is less, than the number of
equations, and one may have many skeletons $\mathcal{Q}^{n},\mathcal{Q}%
^{\prime n},...$placed at the point $Q_{0}$, which are equivalent to
skeleton $\mathcal{P}^{n}$, but they are not equivalent between themselves.
This property is a property of multivariance of the space-time geometry.
This property is actual for simple skeletons, which contain less, than four
points ($n<3$). For instance, for the skeleton of two points $\left\{
P_{0},P_{1}\right\} $, which is described by the vector $\mathbf{P}_{0}%
\mathbf{P}_{1}$, the problem of multivariance is actual. In the space-time
of Minkowski the equivalence of two vectors ($\mathbf{P}_{0}\mathbf{P}_{1}%
\mathrm{eqv}\mathbf{Q}_{0}\mathbf{Q}_{1}$) is single-variant for the
timelike vectors, however it is multivariant for spacelike vectors. In the
general space-time the equivalence relation $\mathbf{P}_{0}\mathbf{P}_{1}%
\mathrm{eqv}\mathbf{Q}_{0}\mathbf{Q}_{1}$ is multivariant for both timelike
and spacelike vectors.

The problem of multivariance is essential for both existence and dynamics of
elementary geometrical objects (elementary particles). Let us formulate
dynamics of elementary particles in the coordinateless form. Dynamics of an
elementary particle, having skeleton $\mathcal{P}^{n}=\left\{
P_{0},P_{1},...P_{n}\right\} $, is described by the world chain 
\begin{eqnarray}
\mathcal{C} &=&\dbigcup\limits_{k}\mathcal{P}_{\left( k\right) }^{n},\qquad 
\mathcal{P}_{\left( s\right) }^{n}=\left\{ P_{0}^{\left( s\right)
},P_{1}^{\left( s\right) },...P_{n}^{\left( s\right) }\right\} ,\qquad 
\mathcal{P}_{\left( 0\right) }^{n}=\mathcal{P}^{n},  \label{c1.7} \\
P_{0}^{\left( s+1\right) } &=&P_{1}^{\left( s\right) }\qquad s=...0,1,2,...
\label{c1.8}
\end{eqnarray}%
Direction of evolution in the space-time is described by the leading vector $%
\mathbf{P}_{0}\mathbf{P}_{1}$. If the motion of the elementary particle is
free, the adjacent links $\mathcal{P}_{\left( s\right) }^{n}$ and $\mathcal{P%
}_{\left( s+1\right) }^{n}$ are equivalent in the sense that 
\begin{equation}
\mathcal{P}_{\left( s\right) }^{n}\mathrm{eqv}\mathcal{P}_{\left( s+1\right)
}^{n}:\qquad \mathbf{P}_{i}^{\left( s\right) }\mathbf{P}_{k}^{\left(
s\right) }\mathrm{eqv}\mathbf{P}_{i}^{\left( s+1\right) }\mathbf{P}%
_{k}^{\left( s+1\right) },\qquad i,k=0,1,...n,\qquad s=...0,1,2,...
\label{c1.9}
\end{equation}

Relations (\ref{c1.7}) - (\ref{c1.9}) realizes coordinateless description of
the free elementary particle motion. In the simplest case, when the
space-time is the space-time of Minkowski, and the skeleton consists of two
points $P_{0},P_{1}$ with timelike leading vector $\mathbf{P}_{0}\mathbf{P}%
_{1}$, the coordinateless description by means of relations (\ref{c1.7}) - (%
\ref{c1.9}) coincides with the conventional description. The conventional
classical dynamics is well defined only in the Riemannian space-time. The
coordinateless dynamic description (\ref{c1.7}) - (\ref{c1.9}) of elementary
particles is a generalization of the conventional classical dynamics onto
the case of arbitrary space-time geometry.

\section{Representations of the proper Euclidean \newline
geometry}

Any geometry is constructed as a modification of the proper Euclidean
geometry. But not all representations of the proper Euclidean geometry are
convenient for modification. There are three representation of the proper
Euclidean geometry \cite{R2007b}. They differ in the number of primary
(basic) elements, forming the Euclidean geometry.

The Euclidean representation (E-representation) contains three basic
elements (point, segment, angle). Any geometrical object (figure) can be
constructed of these basic elements. Properties of the basic elements and
the method of their application are described by the Euclidean axioms.

The vector representation (V-representation) of the proper Euclidean
geometry contains two basic elements (point, vector). The angle is a
derivative element, which is constructed of two vectors. A use of the two
basic elements at the construction of geometrical objects is determined by
the special structure, known as the linear vector space with the scalar
product, given on it (Euclidean space). The scalar product of linear vector
space describes interrelation of two basic elements (vectors), whereas other
properties of the linear vector space associate with the displacement of
vectors.

The third representation ($\sigma $-representation) of the proper Euclidean
geometry contains only one basic element (point). Segment (vector) is a
derivative element. It is constructed of points. The angle is also a
derivative element. It is constructed of two segments (vectors). The $\sigma 
$-representation contains a special structure: world function $\sigma $,
which describes interrelation of two basic elements (points). The world
function $\sigma \left( P_{0},P_{1}\right) =\frac{1}{2}\rho ^{2}\left(
P_{0},P_{1}\right) $, where $\rho \left( P_{0},P_{1}\right) $ is the
distance between points $P_{0}$ and $P_{1}$. The concept of distance $\rho $%
, as well as the world function $\sigma $, is used in all representations of
the proper Euclidean geometry. However, the world function forms a structure
only in the $\sigma $-representation, where the world function $\sigma $
describes interrelation of two basic elements (points). Besides, the world
function satisfies a series of constraints, formulated in terms of $\sigma $
and only in terms of $\sigma $. These conditions (the Euclideaness
conditions) will be formulated below.

The Euclideaness conditions are equivalent to a use of the vector linear
space with the scalar product on it, but formally they do not mention the
linear vector space, because all concepts of the linear vector space, as
well as all concepts of the proper Euclidean geometry are expressed directly
via world function $\sigma $ and only via it.

If we want to modify the proper Euclidean geometry, then we should use the $%
\sigma $-representation for its modification. In the $\sigma $%
-representation the special geometric structure (world function) has the
form of a function of two points. Modifying the form of the world function,
we modify automatically all concepts of the proper Euclidean geometry, which
are expressed via the world function. It is very important, that the
expression of geometrical concepts via the world function does not refer to
the means of description (dimension, coordinate system, concept of a curve).
The fact, that modifying the world function, one violates the Euclideaness
conditions, is of no importance, because one obtains non-Euclidean geometry
as a result of such a modification. A change of the world function means a
change of the distance, which is interpreted as a deformation of the proper
Euclidean geometry. The generalized geometry, obtained by a deformation of
the proper Euclidean geometry is called the tubular geometry (T-geometry),
because in the generalized geometry straight lines are tubes (surfaces), in
general, but not one-dimensional lines. Another name of T-geometry is the
physical geometry. The physical geometry is the geometry, described
completely by the world function. Any physical geometry may be used as a
space-time geometry in the sense, that the set of all T-geometries is the
set of all possible space-time geometries.

Modification of the proper Euclidean geometry in V-representation is very
restricted, because in this representation there are two basic elements.
They are not independent, and one cannot modify them independently. Formally
it means, that the linear vector space is to be preserved as a geometrical
structure. It means, in particular, that the generalized geometry retains to
be continuous, uniform and isotropic. The dimension of the generalized
geometry is to be fixed. Besides, the generalized geometry,obtained by such
a way, cannot be multivariant. Such a property of the space-time geometry as
multivariance can be obtained only in $\sigma $-representation. As far as
the $\sigma $-representation of the proper Euclidean geometry was not known
in the twentieth century, the multivariance of geometry was also unknown
concept.

Transition from the V-representation to $\sigma $-representation is carried
out as follows. All concepts of the linear vector space are expressed in
terms of the world function $\sigma $. In reality, concepts of vector,
scalar product of two vectors and linear dependence of $n$ vectors are
expressed via the world function $\sigma _{\mathrm{E}}$ of the proper
Euclidean geometry. Such operations under vectors as equality of vectors,
summation of vectors and multiplication of a vector by a real number are
expressed by means of some formulae. The characteristic properties of these
operations, which are given in V-representation by means of axioms, are
given now by special properties of the Euclidean world function $\sigma _{%
\mathrm{E}}$. After expression of the linear vector space via the world
function the linear vector space may be not mentioned, because all its
properties are described by the world function. We obtain the $\sigma $%
-representation of the proper Euclidean geometry, where some properties of
the linear vector space are expressed in the form of formulae, whereas
another part of properties is hidden in the specific form of the Euclidean
world function $\sigma _{\mathrm{E}}$. Modifying world function, we modify
automatically the properties of the linear vector space (which is not
mentioned in fact). At such a modification we are not to think about the way
of modification of the linear vector space, which is the principal
geometrical structure in the V-representation. In the $\sigma $%
-representation the linear vector space is a derivative structure, which may
be not mentioned at all. Thus, at transition to $\sigma $-representation the
concepts of the linear vector space (primary concepts in V-representation)
become to be secondary concepts (derivative concepts of the $\sigma $%
-representation).

In $\sigma $-representation we have the following expressions for concepts
of the proper Euclidean geometry. Vector $\mathbf{PQ}=\overrightarrow{PQ}$
is an ordered set of two points $P$ and $Q$. The length $\left\vert \mathbf{%
PQ}\right\vert $ of the vector $\mathbf{PQ}$ is defined by the relation%
\begin{equation}
\left\vert \mathbf{P}_{0}\mathbf{P}_{1}\right\vert =\sqrt{2\sigma \left(
P_{0},P_{1}\right) }  \label{a1.0}
\end{equation}%
The scalar product $\left( \mathbf{P}_{0}\mathbf{P}_{1}.\mathbf{Q}_{0}%
\mathbf{Q}_{1}\right) $ of two vectors $\mathbf{P}_{0}\mathbf{P}_{1}$ and $%
\mathbf{Q}_{0}\mathbf{Q}_{1}$ is defined by the relation%
\begin{equation}
\left( \mathbf{P}_{0}\mathbf{P}_{1}.\mathbf{Q}_{0}\mathbf{Q}_{1}\right)
=\sigma \left( P_{0},Q_{1}\right) +\sigma \left( P_{1},Q_{0}\right) -\sigma
\left( P_{0},Q_{0}\right) -\sigma \left( P_{1},Q_{1}\right)  \label{a1.1}
\end{equation}%
where the world function $\sigma $%
\begin{equation}
\sigma :\qquad \Omega \times \Omega \rightarrow \mathbb{R},\qquad \sigma
\left( P,Q\right) =\sigma \left( Q,P\right) ,\qquad \sigma \left( P,P\right)
=0,\qquad \forall P,Q\in \Omega  \label{a1.2}
\end{equation}%
is the world function $\sigma _{\mathrm{E}}$ of the Euclidean geometry.

In the proper Euclidean geometry $n$ vectors $\mathbf{P}_{0}\mathbf{P}_{k}$, 
$k=1,2,...n$ are linear dependent, if and only if the Gram's determinant 
\begin{equation}
F\left( \mathcal{P}^{n}\right) =0,\qquad \mathcal{P}^{n}=\left\{
P_{0},P_{1},...,P_{n}\right\}  \label{a1.2a}
\end{equation}
where the Gram's determinant $F\left( \mathcal{P}^{n}\right) $ is defined by
the relation 
\begin{equation}
F\left( \mathcal{P}^{n}\right) \equiv \det \left\vert |\left( \mathbf{P}_{0}%
\mathbf{P}_{i}.\mathbf{P}_{0}\mathbf{P}_{k}\right) |\right\vert ,\qquad
i,k=1,2,...n  \label{a1.7}
\end{equation}%
Using expression (\ref{a1.1}) for the scalar product, the condition of the
linear dependence of $n$ vectors $\mathbf{P}_{0}\mathbf{P}_{k}$, $k=1,2,...n$
is written in the form%
\begin{equation}
F\left( \mathcal{P}^{n}\right) \equiv \det \left\vert |\sigma \left(
P_{0},P_{i}\right) +\sigma \left( P_{0},P_{k}\right) -\sigma \left(
P_{i},P_{k}\right) |\right\vert =0,\qquad i,k=1,2,...n  \label{a1.7b}
\end{equation}

Definition (\ref{a1.1}) of the scalar product of two vectors coincides with
the conventional scalar product of vectors in the proper Euclidean space.
(One can verify this easily). The relations (\ref{a1.1}), (\ref{a1.7b}) do
not contain a reference to the dimension of the Euclidean space and to a
coordinate system in it. Hence, the relations (\ref{a1.1}), (\ref{a1.7b})
are general geometric relations, which may be considered as a definition of
the scalar product of two vectors and that of the linear dependence of
vectors.

Equivalence (equality) of two vectors $\mathbf{P}_{0}\mathbf{P}_{1}$ and $%
\mathbf{Q}_{0}\mathbf{Q}_{1}$ is defined by the relations%
\begin{equation}
\mathbf{P}_{0}\mathbf{P}_{1}\text{eqv}\mathbf{Q}_{0}\mathbf{Q}_{1}:\qquad
\left( \mathbf{P}_{0}\mathbf{P}_{1}.\mathbf{Q}_{0}\mathbf{Q}_{1}\right)
=\left\vert \mathbf{P}_{0}\mathbf{P}_{1}\right\vert \cdot \left\vert \mathbf{%
Q}_{0}\mathbf{Q}_{1}\right\vert \wedge \left\vert \mathbf{P}_{0}\mathbf{P}%
_{1}\right\vert =\left\vert \mathbf{Q}_{0}\mathbf{Q}_{1}\right\vert
\label{a1.3}
\end{equation}%
where $\left\vert \mathbf{P}_{0}\mathbf{P}_{1}\right\vert $ is the length (%
\ref{a1.0}) of the vector $\mathbf{P}_{0}\mathbf{P}_{1}$%
\begin{equation}
\left\vert \mathbf{P}_{0}\mathbf{P}_{1}\right\vert =\sqrt{\left( \mathbf{P}%
_{0}\mathbf{P}_{1}.\mathbf{P}_{0}\mathbf{P}_{1}\right) }=\sqrt{2\sigma
\left( P_{0},P_{1}\right) }  \label{a1.4}
\end{equation}

In the developed form the condition (\ref{a1.3}) of equivalence of two
vectors $\mathbf{P}_{0}\mathbf{P}_{1}$ and $\mathbf{Q}_{0}\mathbf{Q}_{1}$
has the form%
\begin{eqnarray}
\sigma \left( P_{0},Q_{1}\right) +\sigma \left( P_{1},Q_{0}\right) -\sigma
\left( P_{0},Q_{0}\right) -\sigma \left( P_{1},Q_{1}\right) &=&2\sigma
\left( P_{0},P_{1}\right)  \label{a1.5} \\
\sigma \left( P_{0},P_{1}\right) &=&\sigma \left( Q_{0},Q_{1}\right)
\label{a1.6}
\end{eqnarray}

Let the points $P_{0},P_{1}$, determining the vector $\mathbf{P}_{0}\mathbf{P%
}_{1}$, and the origin $Q_{0}$ of the vector $\mathbf{Q}_{0}\mathbf{Q}_{1}$
be given. Let $\mathbf{P}_{0}\mathbf{P}_{1}$eqv$\mathbf{Q}_{0}\mathbf{Q}_{1}$%
. We can determine the vector $\mathbf{Q}_{0}\mathbf{Q}_{1}$, solving two
equations (\ref{a1.5}), (\ref{a1.6}) with respect to the position of the
point $Q_{1}$.

In the case of the proper Euclidean space there is one and only one solution
of equations (\ref{a1.5}), (\ref{a1.6}) independently of the space dimension 
$n$. In the case of arbitrary T-geometry one can guarantee neither existence
nor uniqueness of the solution of equations (\ref{a1.5}), (\ref{a1.6}) for
the point $Q_{1}$. Number of solutions depends on the form of the world
function $\sigma $. This fact means a multivariance of the property of two
vectors equivalence in the arbitrary T-geometry. In other words, the
single-variance of the vector equality in the proper Euclidean space is a
specific property of the proper Euclidean geometry, and this property is
conditioned by the form of the Euclidean world function. In other
T-geometries this property does not take place, in general.

The multivariance is a general property of a physical geometry. It is
connected with a necessity of solution of algebraic equations, containing
the world function. As far as the world function is different in different
physical geometries, the solution of these equations may be not unique, or
it may not exist at all.

If in the $n$-dimensional Euclidean space $F\left( \mathcal{P}^{n}\right)
\neq 0$, the vectors $\mathbf{P}_{0}\mathbf{P}_{k}$, $k=1,2,...n$ are linear
independent. We may construct rectilinear coordinate system with basic
vectors $\mathbf{P}_{0}\mathbf{P}_{k}$, $k=1,2,...n$ in the $n$-dimensional
Euclidean space. Covariant coordinates $x_{k}=\left( \mathbf{P}_{0}\mathbf{P}%
\right) _{k}$ of the vector $\mathbf{P}_{0}\mathbf{P}$ in this coordinate
system have the form%
\begin{equation}
x_{k}=x_{k}\left( P\right) =\left( \mathbf{P}_{0}\mathbf{P}\right)
_{k}=\left( \mathbf{P}_{0}\mathbf{P}.\mathbf{P}_{0}\mathbf{P}_{k}\right)
,\qquad k=1,2,...n  \label{a1.8}
\end{equation}

Now we can formulate the Euclideaness conditions. These conditions are
conditions of the fact, that the T-geometry, described by the world function 
$\sigma $, is $n$-dimensional proper Euclidean geometry.

I. Definition of the dimension and introduction of the rectilinear
coordinate system: 
\begin{equation}
\exists \mathcal{P}^{n}\equiv \left\{ P_{0},P_{1},...P_{n}\right\} \subset
\Omega ,\qquad F_{n}\left( \mathcal{P}^{n}\right) \neq 0,\qquad F_{k}\left( {%
\Omega }^{k+1}\right) =0,\qquad k>n  \label{g2.5}
\end{equation}%
where $F_{n}\left( \mathcal{P}^{n}\right) $\ is the Gram's determinant (\ref%
{a1.7}). Vectors $\mathbf{P}_{0}\mathbf{P}_{i}$, $\;i=1,2,...n$\ are basic
vectors of the rectilinear coordinate system $K_{n}$\ with the origin at the
point $P_{0}$. In $K_{n}$ the covariant metric tensor $g_{ik}\left( \mathcal{%
P}^{n}\right) $, \ $i,k=1,2,...n$\ and the contravariant one $g^{ik}\left( 
\mathcal{P}^{n}\right) $, \ $i,k=1,2,...n$\ \ are defined by the relations 
\begin{equation}
\sum\limits_{k=1}^{k=n}g^{ik}\left( \mathcal{P}^{n}\right) g_{lk}\left( 
\mathcal{P}^{n}\right) =\delta _{l}^{i},\qquad g_{il}\left( \mathcal{P}%
^{n}\right) =\left( \mathbf{P}_{0}\mathbf{P}_{i}.\mathbf{P}_{0}\mathbf{P}%
_{l}\right) ,\qquad i,l=1,2,...n  \label{a1.5b}
\end{equation}%
\begin{equation}
F_{n}\left( \mathcal{P}^{n}\right) =\det \left\vert \left\vert g_{ik}\left( 
\mathcal{P}^{n}\right) \right\vert \right\vert \neq 0,\qquad i,k=1,2,...n
\label{g2.6}
\end{equation}

II. Linear structure of the Euclidean space: 
\begin{equation}
\sigma \left( P,Q\right) =\frac{1}{2}\sum\limits_{i,k=1}^{i,k=n}g^{ik}\left( 
\mathcal{P}^{n}\right) \left( x_{i}\left( P\right) -x_{i}\left( Q\right)
\right) \left( x_{k}\left( P\right) -x_{k}\left( Q\right) \right) ,\qquad
\forall P,Q\in \Omega  \label{a1.5a}
\end{equation}%
where coordinates $x_{i}=x_{i}\left( P\right) ,$\ $i=1,2,...n$\ of the point 
$P$\ are covariant coordinates of the vector $\mathbf{P}_{0}\mathbf{P}$,
defined by the relation (\ref{a1.8}).

III: The metric tensor matrix $g_{lk}\left( \mathcal{P}^{n}\right) $\ has
only positive eigenvalues 
\begin{equation}
g_{k}>0,\qquad k=1,2,...,n  \label{a15c}
\end{equation}

IV. The continuity condition: the system of equations 
\begin{equation}
\left( \mathbf{P}_{0}\mathbf{P}_{i}.\mathbf{P}_{0}\mathbf{P}\right)
=y_{i}\in \mathbb{R},\qquad i=1,2,...n  \label{b14}
\end{equation}%
considered to be equations for determination of the point $P$\ as a function
of coordinates $y=\left\{ y_{i}\right\} $,\ \ $i=1,2,...n$\ has always one
and only one solution.\textit{\ }All conditions I $\div $ IV contain a
reference to the dimension $n$\ of the Euclidean space.

One can show that conditions I $\div $ IV are the necessary and sufficient
conditions of the fact that the set $\Omega $ together with the world
function $\sigma $, given on $\Omega \times \Omega $, describes the $n$%
-dimensional Euclidean space \cite{R90}.

Investigation of the Dirac particle (dynamic system, described by the Dirac
equation) has shown, that the Dirac particle is a composite particle \cite%
{R2004}, whose internal degrees of freedom are described nonrelativistically 
\cite{R2004b}. The composite structure of the Dirac particle may be
explained as a relativistic rotator, consisting of two (or more) particles,
rotating around their inertia centre. The relativistic rotator explains
existence of the Dirac particle spin, however, the problem of the rotating
particles confinement appears. In this paper we try to explain the problem
of spin in the framework of the program of the physics geometrization, when
dynamics of physical bodies is determined by the space-time geometry.

Although the first stages of the physics geometrization (the special
relativity and the general relativity) manifest themselves very well, the
papers on further geometrization of physics, which ignore the quantum
principles, are considered usually as dissident.

\section{Dynamics as a result of the space-time \newline
geometry}

Dynamics in the space-time, described by a physical geometry (T-geometry),
is presented in \cite{R2007a}. Here we remind the statement of the problem
of dynamics.

Geometrical object $\mathcal{O\subset }\Omega $ is a subset of points in the
point set $\Omega $. In the T-geometry the geometric object $\mathcal{O}$ is
described by means of the skeleton-envelope method. It means that any
geometric object $\mathcal{O}$ is considered to be a set of intersections
and joins of elementary geometric objects (EGO).

The elementary geometrical object $\mathcal{E}$ is described by its skeleton 
$\mathcal{P}^{n}$ and envelope function $f_{\mathcal{P}^{n}}$. The finite
set $\mathcal{P}^{n}\equiv \left\{ P_{0},P_{1},...,P_{n}\right\} \subset
\Omega $ of parameters of the envelope function $f_{\mathcal{P}^{n}}$ is the
skeleton of elementary geometric object (EGO) $\mathcal{E}\subset \Omega $.
The set $\mathcal{E}\subset \Omega $ of points forming EGO is called the
envelope of its skeleton $\mathcal{P}^{n}$. The envelope function $f_{%
\mathcal{P}^{n}}$%
\begin{equation}
f_{\mathcal{P}^{n}}:\qquad \Omega \rightarrow \mathbb{R},  \label{h2.1}
\end{equation}%
determining EGO is a function of the running point $R\in \Omega $ and of
parameters $\mathcal{P}^{n}\subset \Omega $. The envelope function $f_{%
\mathcal{P}^{n}}$ is supposed to be an algebraic function of $s$ arguments $%
w=\left\{ w_{1},w_{2},...w_{s}\right\} $, $s=(n+2)(n+1)/2$. Each of
arguments $w_{k}=\sigma \left( Q_{k},L_{k}\right) $ is the world function $%
\sigma $ of two points $Q_{k},L_{k}\in \left\{ R,\mathcal{P}^{n}\right\} $,
either belonging to skeleton $\mathcal{P}^{n}$, or coinciding with the
running point $R$. Thus, any elementary geometric object $\mathcal{E}$ is
determined by its skeleton $\mathcal{P}^{n}$ and its envelope function $f_{%
\mathcal{P}^{n}}$. Elementary geometric object $\mathcal{E}$ is the set of
zeros of the envelope function 
\begin{equation}
\mathcal{E}=\left\{ R|f_{\mathcal{P}^{n}}\left( R\right) =0\right\}
\label{h2.2}
\end{equation}%
\textit{Definition.} Two EGOs $\mathcal{E}_{\mathcal{P}^{n}}$ and $\mathcal{E%
}_{\mathcal{Q}^{n}}$ are equivalent, if their skeletons $\mathcal{P}^{n}$
and $\mathcal{Q}^{n}$ are equivalent and their envelope functions $f_{%
\mathcal{P}^{n}}$ and $g_{\mathcal{Q}^{n}}$ are equivalent. Equivalence ($%
\mathcal{P}^{n}$eqv$\mathcal{Q}^{n}$) of two skeletons $\mathcal{P}%
^{n}\equiv \left\{ P_{0},P_{1},...,P_{n}\right\} \subset \Omega $ and $%
\mathcal{Q}^{n}\equiv \left\{ Q_{0},Q_{1},...,Q_{n}\right\} \subset \Omega $
means that 
\begin{equation}
\mathcal{P}^{n}\text{eqv}\mathcal{Q}^{n}:\qquad \mathbf{P}_{i}\mathbf{P}_{k}%
\text{eqv}\mathbf{Q}_{i}\mathbf{Q}_{k},\qquad i,k=0,1,...n,\quad i\leq k
\label{a5.4}
\end{equation}%
Equivalence of the envelope functions $f_{\mathcal{P}^{n}}$ and $g_{\mathcal{%
Q}^{n}}$ means, that they have the same set of zeros. It means that 
\begin{equation}
f_{\mathcal{P}^{n}}\left( R\right) =\Phi \left( g_{\mathcal{P}^{n}}\left(
R\right) \right) ,\qquad \forall R\in \Omega  \label{a5.5}
\end{equation}%
where $\Phi $ is an arbitrary function, having the property%
\begin{equation}
\Phi :\mathbb{R}\rightarrow \mathbb{R},\qquad \Phi \left( 0\right) =0
\label{a5.5a}
\end{equation}

Evolution of EGO $\mathcal{O}_{\mathcal{P}^{n}}$ in the space-time is
described as a world chain $\mathcal{C}_{\mathrm{fr}}$ of equivalent
connected EGOs. The point $P_{0}$ of the skeleton $\mathcal{P}^{n}=\left\{
P_{0},P_{1},...P_{n}\right\} $ is considered to be the origin of the
geometrical object $\mathcal{O}_{\mathcal{P}^{n}}.$ The EGO $\mathcal{O}_{%
\mathcal{P}^{n}}$ is considered to be placed at its origin $P_{0}$. Let us
consider a set of equivalent skeletons $\mathcal{P}_{\left( l\right)
}^{n}=\left\{ P_{0}^{\left( l\right) },P_{1}^{\left( l\right)
},...P_{n}^{\left( l\right) }\right\} ,$ $l=...0,1,...$which are equivalent
in pairs 
\begin{equation}
\mathbf{P}_{i}^{\left( l\right) }\mathbf{P}_{k}^{\left( l\right) }\text{eqv}%
\mathbf{P}_{i}^{\left( l+1\right) }\mathbf{P}_{k}^{\left( l+1\right)
},\qquad i,k=0,1,...n;\qquad l=...1,2,...  \label{a6.1}
\end{equation}%
The skeletons $\mathcal{P}_{\left( l\right) }^{n},$ $l=...0,1,...$are
connected, and they form a chain in the direction of vector $\mathbf{P}_{0}%
\mathbf{P}_{1}$, if the point $P_{1}$ of one skeleton coincides with the
origin $P_{0}$ of the adjacent skeleton 
\begin{equation}
P_{1}^{\left( l\right) }=P_{0}^{\left( l+1\right) },\qquad l=...0,1,2,...
\label{a6.2}
\end{equation}%
The chain $\mathcal{C}_{\mathrm{fr}}$ describes evolution of the elementary
geometrical object $\mathcal{O}_{\mathcal{P}^{n}}$ in the direction of the
leading vector $\mathbf{P}_{0}\mathbf{P}_{1}$. The evolution of EGO $%
\mathcal{O}_{\mathcal{P}^{n}}$ is a temporal evolution, if the vectors $%
\mathbf{P}_{0}^{\left( l\right) }\mathbf{P}_{1}^{\left( l\right) }$ are
timelike $\left\vert \mathbf{P}_{0}^{\left( l\right) }\mathbf{P}_{1}^{\left(
l\right) }\right\vert ^{2}>0,$ $\ l=...0,1,..$. The evolution of EGO $%
\mathcal{O}_{\mathcal{P}^{n}}$ is a spatial evolution, if the vectors $%
\mathbf{P}_{0}^{\left( l\right) }\mathbf{P}_{1}^{\left( l\right) }$ are
spacelike $\left\vert \mathbf{P}_{0}^{\left( l\right) }\mathbf{P}%
_{1}^{\left( l\right) }\right\vert ^{2}<0,$ $\ l=...0,1,..$.

Note, that all adjacent links (EGOs) of the chain are equivalent in pairs,
although two links of the chain may be not equivalent, if they are not
adjacent. However, lengths of corresponding vectors are equal in all links
of the chain 
\begin{equation}
\left\vert \mathbf{P}_{i}^{\left( l\right) }\mathbf{P}_{k}^{\left( l\right)
}\right\vert =\left\vert \mathbf{P}_{i}^{\left( s\right) }\mathbf{P}%
_{k}^{\left( s\right) }\right\vert ,\qquad i,k=0,1,...n;\qquad l,s=...1,2,...
\label{a6.4}
\end{equation}%
We shall refer to the vector $\mathbf{P}_{0}^{\left( l\right) }\mathbf{P}%
_{1}^{\left( l\right) }$, which determines the form of the evolution and the
shape of the world chain, as the leading vector. This vector determines the
direction of 4-velocity of the physical body, associated with the link of
the world chain.

If the relations%
\begin{eqnarray}
\mathcal{P}^{n}\mathrm{eqv}\mathcal{Q}^{n} &:&\quad \left( \mathbf{P}_{i}%
\mathbf{P}_{k}.\mathbf{Q}_{i}\mathbf{Q}_{k}\right) =\left\vert \mathbf{P}_{i}%
\mathbf{P}_{k}\right\vert \cdot \left\vert \mathbf{Q}_{i}\mathbf{Q}%
_{k}\right\vert ,\qquad \left\vert \mathbf{P}_{i}\mathbf{P}_{k}\right\vert
=\left\vert \mathbf{Q}_{i}\mathbf{Q}_{k}\right\vert ,  \label{a6.9} \\
i,k &=&0,1,2,...n
\end{eqnarray}%
\begin{eqnarray}
\mathcal{Q}^{n}\mathrm{eqv}\mathcal{R}^{n} &:&\quad \left( \mathbf{Q}_{i}%
\mathbf{Q}_{k}.\mathbf{R}_{i}\mathbf{R}_{k}\right) =\left\vert \mathbf{Q}_{i}%
\mathbf{Q}_{k}\right\vert \cdot \left\vert \mathbf{R}_{i}\mathbf{R}%
_{k}\right\vert ,\qquad \left\vert \mathbf{Q}_{i}\mathbf{Q}_{k}\right\vert
=\left\vert \mathbf{R}_{i}\mathbf{R}_{k}\right\vert ,  \label{a6.10} \\
i,k &=&0,1,2,...n
\end{eqnarray}%
are satisfied, the relations%
\begin{eqnarray}
\mathcal{P}^{n}\mathrm{eqv}\mathcal{R}^{n} &:&\quad \left( \mathbf{P}_{i}%
\mathbf{P}_{k}.\mathbf{R}_{i}\mathbf{R}_{k}\right) =\left\vert \mathbf{P}_{i}%
\mathbf{P}_{k}\right\vert \cdot \left\vert \mathbf{R}_{i}\mathbf{R}%
_{k}\right\vert ,\qquad \left\vert \mathbf{P}_{i}\mathbf{P}_{k}\right\vert
=\left\vert \mathbf{R}_{i}\mathbf{R}_{k}\right\vert ,  \label{a6.11} \\
i,k &=&0,1,2,...n
\end{eqnarray}%
are not satisfied, in general, because the relations (\ref{a6.11}) contain
the scalar products $\left( \mathbf{P}_{i}\mathbf{P}_{k}.\mathbf{R}_{i}%
\mathbf{R}_{k}\right) $. These scalar products contain the world functions $%
\sigma \left( P_{i},R_{k}\right) $, which are not contained in relations (%
\ref{a6.9}), (\ref{a6.10}).

The world chain $\mathcal{C}_{\mathrm{fr}}$, consisting of equivalent links (%
\ref{a6.1}), (\ref{a6.2}), describes a free motion of a physical body
(particle), associated with the skeleton $\mathcal{P}^{n}$. We assume that 
\textit{the motion of physical body is free, if all points of the body move
free} (i.e. without acceleration). If the external forces are absent, the
physical body as a whole moves without acceleration. However, if the body
rotates, one may not consider a motion of this body as a free motion,
because not all points of this body move free (without acceleration). In the
rotating body there are internal forces, which generate centripetal
acceleration of some points of the body. As a result some points of the body
do not move free. Motion of the rotating body may be free only on the
average, but not exactly free.

Conception of non-free motion of a particle is rather indefinite, and we
restrict ourselves with consideration of a free motion only.

Conventional conception of the motion of extensive (non-pointlike) particle,
which is free on the average, contains a free displacement, described by the
velocity 4-vector, and a spatial rotation, described by the angular velocity
3-pseudovector $\mathbf{\omega }$. The velocity 4-vector is associated with
the timelike leading vector $\mathbf{P}_{0}\mathbf{P}_{1}$. At the free on
the average motion of a rotating body some of vectors $\mathbf{P}_{0}\mathbf{%
P}_{2}^{\mathrm{(s)}},\mathbf{P}_{0}\mathbf{P}_{3}^{\mathrm{(s)}},$...of the
skeleton $\mathcal{P}^{n}$ are not in parallel with vectors $\mathbf{P}_{0}%
\mathbf{P}_{2}^{\mathrm{(s+1)}},\mathbf{P}_{0}\mathbf{P}_{3}^{\mathrm{(s+1)}%
},...$, although at the free motion all vectors $\mathbf{P}_{0}\mathbf{P}%
_{2}^{\mathrm{(s)}},\mathbf{P}_{0}\mathbf{P}_{3}^{\mathrm{(s)}},...$ are to
be in parallel with $\mathbf{P}_{0}\mathbf{P}_{2}^{\mathrm{(s+1)}},\mathbf{P}%
_{0}\mathbf{P}_{3}^{\mathrm{(s+1)}},...$ as follows from (\ref{a6.1}). It
means that the world chain $\mathcal{C}_{\mathrm{fr}}$ of a freely moving
body can describe only translation of a physical body, but not its rotation.

If the leading vector $\mathbf{P}_{0}\mathbf{P}_{1}$ is spacelike, the body,
described by the skeleton $\mathcal{P}^{n}$, evolves in the spacelike
direction. It seems, that the spacelike evolution is prohibited. But it is
not so. If the world chain forms a helix with the timelike axis, such a
world chain may be considered as timelike on the average. In reality such
world chains are possible. For instance, the world chain of the classical
Dirac particle is a helix with timelike axis. It is not quite clear, whether
or not the links of this chain are spacelike, because internal degrees of
freedom of the Dirac particle, responsible for helicity of the world chain,
are described nonrelativistically.

Thus, consideration of a spatial evolution is not meaningless, especially if
we take into account, that the spatial evolution may imitate rotation, which
is absent at the free motion of a particle. Further we consider the problem
of the spatial evolution.

\section{Dynamics of classical Dirac particle}

Dirac particle $\mathcal{S}_{\mathrm{D}}$ is the dynamic system, described
by the Dirac equation. The free Dirac particle $\mathcal{S}_{\mathrm{D}}$ is
described by the free Dirac equation 
\begin{equation}
i\hbar \gamma ^{l}\partial _{l}\psi -m\psi =0  \label{f1.2}
\end{equation}%
where $\psi $ is the four-component complex wave function, and $\gamma ^{l}$%
, $l=0,1,2,3$ are $4\times 4$ complex matrices, satisfying the relations%
\begin{equation*}
\gamma ^{i}\gamma ^{k}+\gamma ^{k}\gamma ^{i}=2Ig^{ik},\qquad i,k=0,1,2,3,
\end{equation*}%
$I$ is the $4\times 4$ unit matrix, $g^{ik}$ is the metric tensor.
Expressions of physical quantities: the 4-flux $j^{k}$ of particles and the
energy-momentum tensor $T_{l}^{k}$ have the form%
\begin{equation}
j^{k}=\bar{\psi}\gamma ^{k}\psi ,\qquad T_{l}^{k}=\frac{i}{2}\left( \bar{\psi%
}\gamma ^{k}\partial _{l}\psi -\partial _{l}\bar{\psi}\cdot \gamma ^{k}\psi
\right) ,\qquad k,l=0,1,2,3  \label{f1.3}
\end{equation}%
where $\bar{\psi}=\psi ^{\ast }\gamma ^{0}$, $\psi ^{\ast }$ is the
Hermitian conjugate to $\psi $. The classical Dirac particle is a dynamic
system $\mathcal{S}_{\mathrm{Dcl}}$, which is obtained from the dynamic
system $\mathcal{S}_{\mathrm{D}}$ in the classical limit.

To obtain the classical limit, one may not set the quantum constant $\hbar
=0 $ in the equation (\ref{f1.2}), because in this case we do not obtain any
reasonable description of the particle.

The Dirac particle $\mathcal{S}_{\mathrm{D}}$ is a quantum particle in the
sense, that it is described by a system of partial differential equations
(PDE), which contain the quantum constant $\hbar $. The classical Dirac
particle $\mathcal{S}_{\mathrm{Dcl}}$ is described by a system of ordinary
differential equations (ODE), which contain the quantum constant $\hbar $ as
a parameter. May the system of ODE carry out the classical description, if
it contains the quantum constant $\hbar $? The answer depends on the
viewpoint of investigator. If the investigator believes that \textit{the
quantum constant is an attribute of quantum principles and only of quantum
principles}, he supposes that, containing $\hbar $, the dynamic equations
cannot realize a classical description, where the quantum principles are not
used. However, if the investigator consider the classical description simply
as method of investigation of the quantum dynamic equations, it is of no
importance, whether or not the system of ODE contains the quantum constant.
It is important only, that the system of PDE is approximated by a system of
ODE. The dynamic system, described by PDE, contains infinite number of the
freedom degrees. The dynamic system, described by ODE, contains several
degrees of freedom. It is simpler and can be investigated more effectively.

Obtaining the classical approximation, we use the procedure of dynamic
disquantization \cite{R2001c}. This procedure transforms the system of PDE
into the system of ODE. The procedure of dynamic disquantization is a
dynamical procedure, which has no relation to the process of quantization or
disquantization in the sense, that it does not refer to the quantum
principles. The dynamic disquantization means that all derivatives $\partial
_{k}$ in dynamic equations are replaced by the projection of vector $%
\partial _{k}$ onto the current vector $j^{k}$ 
\begin{equation}
\partial _{k}\longrightarrow \frac{j_{k}}{j_{l}j^{l}}j^{s}\partial _{s}
\label{b3.1}
\end{equation}%
This dynamical operation is called the dynamic disquantization, because,
applying it to the Schr\"{o}dinger equation, we obtain the dynamic equations
for the statistical ensemble of classical nonrelativistic particles. These
dynamic equations are ODE, which do not depend on the quantum constant $%
\hbar $.

Applying the operation (\ref{b3.1}), to the Dirac equation (\ref{f1.2}), we
transform it to the form%
\begin{equation}
i\hbar \gamma ^{l}\frac{j_{l}}{j^{k}j_{k}}j^{s}\partial _{s}\psi -m\psi
=0,\qquad j^{k}=\bar{\psi}\gamma ^{k}\psi  \label{b3.2}
\end{equation}%
The equation (\ref{b3.2}) is the dynamic equation for the dynamic system
system $\mathcal{E}_{\mathrm{Dqu}}$. The equation (\ref{b3.2}) contains only
derivative $j^{s}\partial _{s}=\left( \bar{\psi}\gamma ^{s}\psi \right)
\partial _{s}$ in the direction of the current 4-vector $j^{k}$. In terms of
the wave function $\psi $ the dynamic equation (\ref{b3.2}) for $\mathcal{E}%
_{\mathrm{Dqu}}$ looks rather bulky. However, in the properly chosen
variables the action for the dynamic system $\mathcal{E}_{\mathrm{Dqu}}$ has
the form \cite{R2001c} 
\begin{equation}
\mathcal{A}_{\mathrm{Dqu}}[x,\mathbf{\xi }]=\int \left\{ -\kappa _{0}m\sqrt{%
\dot{x}^{i}\dot{x}_{i}}+\hbar {\frac{(\dot{\mathbf{\xi }}\times \mathbf{\xi }%
)\mathbf{z}}{2(1+\mathbf{\xi z})}}+\hbar \frac{(\dot{\mathbf{x}}\times \ddot{%
\mathbf{x}})\mathbf{\xi }}{2\sqrt{\dot{x}^{s}\dot{x}_{s}}(\sqrt{\dot{x}^{s}%
\dot{x}_{s}}+\dot{x}^{0})}\right\} d^{4}\tau  \label{a5.18}
\end{equation}%
where the dot means the total derivative $\dot{x}^{s}\equiv dx^{s}/d\tau
_{0} $.\ The quantities $x=\left\{ x^{0},\mathbf{x}\right\} =\{x^{i}\}$, $%
\;i=0,1,2,3$, $\mathbf{\xi }=\{\xi ^{\alpha }\}$, $\alpha =1,2,3$ are
considered to be functions of the Lagrangian coordinates $\tau _{0}$, $%
\mathbf{\tau }=\{\tau _{1},\tau _{2},\tau _{3}\}$. The variables $x$
describe position of the Dirac particle. Here and in what follows the symbol 
$\times $ means the vector product of two 3-vectors. The quantity$\;\mathbf{z%
}$ is the constant unit 3-vector, $\kappa _{0}$ is a dichotomic quantity $%
\kappa _{0}=\pm 1$, $m$ is the constant (mass) taken from the Dirac equation
(\ref{f1.2}). In fact, variables $x$ depend on $\mathbf{\tau }$ as on
parameters, because the action (\ref{a5.18}) does not contain derivatives
with respect to $\tau _{\alpha }$, \ $\alpha =1,2,3$. Lagrangian density of
the action (\ref{a5.18}) does not contain independent variables $\tau $
explicitly. Hence, it may be written in the form 
\begin{equation}
\mathcal{A}_{\mathrm{Dqu}}[x,\mathbf{\xi }]=\int \mathcal{A}_{\mathrm{Dcl}%
}[x,\mathbf{\xi }]d\mathbf{\tau ,\qquad }d\mathbf{\tau }=d\tau _{1}d\tau
_{2}d\tau _{3}  \label{b3.8}
\end{equation}%
where 
\begin{equation}
\mathcal{S}_{\mathrm{Dcl}}:\qquad \mathcal{A}_{\mathrm{Dcl}}[x,\mathbf{\xi }%
]=\int \left\{ -\kappa _{0}m\sqrt{\dot{x}^{i}\dot{x}_{i}}+\hbar {\frac{(\dot{%
\mathbf{\xi }}\times \mathbf{\xi })\mathbf{z}}{2(1+\mathbf{\xi z})}}+\hbar 
\frac{(\dot{\mathbf{x}}\times \ddot{\mathbf{x}})\mathbf{\xi }}{2\sqrt{\dot{x}%
^{s}\dot{x}_{s}}(\sqrt{\dot{x}^{s}\dot{x}_{s}}+\dot{x}^{0})}\right\} d\tau
_{0}  \label{b3.9}
\end{equation}

The action (\ref{b3.8}) is the action for the dynamic system $\mathcal{E}_{%
\mathrm{Dqu}}$, which is a set of similar independent dynamic systems $%
\mathcal{S}_{\mathrm{Dcl}}$. Such a dynamic system is called a statistical
ensemble. Dynamic systems $\mathcal{S}_{\mathrm{Dcl}}$ are elements
(constituents) of the statistical ensemble $\mathcal{E}_{\mathrm{Dqu}}$.
Dynamic equations for each $\mathcal{S}_{\mathrm{Dcl}}$ form a system of
ordinary differential equations. It may be interpreted in the sense, that
the dynamic system $\mathcal{S}_{\mathrm{Dcl}}$ may be considered to be a
classical one, although Lagrangian of $\mathcal{S}_{\mathrm{Dcl}}$ contains
the quantum constant $\hbar $. The dynamic system $\mathcal{S}_{\mathrm{Dcl}%
} $ will be referred to as the classical Dirac particle.

The dynamic system $\mathcal{S}_{\mathrm{Dcl}}$ has ten degrees of freedom.
It describes a composite particle \cite{R2004}. External degrees of freedom
are described relativistically by variables $x$. Internal degrees of freedom
are described nonrelativistically \cite{R2004b} by variables $\mathbf{\xi }$%
. Solution of dynamic equations, generated by the action (\ref{b3.9}), gives
the following result \cite{R2004}. In the coordinate system, where the
canonical momentum four-vector $P_{k}$ has the form 
\begin{equation}
P_{k}=\left\{ p_{0},\mathbf{p}\right\} =\left\{ -\left( 2-\frac{1}{\gamma }%
\right) \kappa _{0}m,0,0,0\right\}  \label{e6.31a}
\end{equation}%
the world line of the classical Dirac particle is a helix, which is
described by the relation 
\begin{eqnarray}
\left\{ t,\mathbf{x}\right\} &=&\left\{ t,R\sin \left( \Omega t\right)
,R\cos \left( \Omega t\right) ,0\right\}  \label{e6.33} \\
R &=&\frac{\hbar \gamma \sqrt{\gamma ^{2}-1}}{2m},\qquad \Omega =\frac{2m}{%
\hbar \gamma ^{2}}  \label{e6.34}
\end{eqnarray}%
where the speed of the light $c=1$, and $\gamma $ is an arbitrary constant
(Lorentz factor of the classical Dirac particle). The velocity $\mathbf{v}=d%
\mathbf{x}/dt$ of the classical Dirac particle is expressed as follows 
\begin{equation}
\mathbf{v}^{2}=1-\frac{1}{\gamma ^{2}},\qquad \gamma =\frac{1}{\sqrt{1-%
\mathbf{v}^{2}}}  \label{e6.32}
\end{equation}

Helical world line of the classical Dirac particle means a rotation of the
particle around some point. On the one hand, such a rotation seems to be
reasonable, because it explains freely the Dirac particle spin and magnetic
moment. On the other hand, the description of this rotation is
nonrelativistic. Besides, it seems rather strange, that the world line of a
free classical particle is a helix, but not a straight line. Attempt of
consideration of the Dirac particle as a rotator, consisting of two
particles \cite{R2004}, meets the problem of confinement of the two
particles.

Although the pure dynamical methods of investigation are more general and
effective, than the investigation methods, based on quantum principles, the
purely dynamical methods of investigation meet incomprehension of most
investigators, who believe, that the Dirac particle must be investigated by
quantum methods. The papers, devoted to investigation of the Dirac equation
by the dynamic methods, are considered as dissident. They are rejected by
the peer review journals (see discussion in \cite{R2005a,R2006}).

Suddenly it is discovered that the helical world line, which is
characteristic for the classical Dirac particle, can be obtained as a result
of a spatial evolution of geometric objects in the framework of properly
chosen space-time geometry.

\section{Existence of such a space-time geometry, where a spatial evolution
may look as world line of classical Dirac particle}

Let us consider the flat homogeneous isotropic space-time $V_{\mathrm{d}%
}=\left\{ \sigma _{\mathrm{d}},\mathbb{R}^{4}\right\} $, described by the
world function 
\begin{equation}
\sigma _{\mathrm{d}}=\sigma _{\mathrm{M}}+d\cdot \mathrm{sgn}\left( \sigma _{%
\mathrm{M}}\right)  \label{a4.0}
\end{equation}%
\begin{equation}
d=\lambda _{0}^{2}=\text{const}>0  \label{a4.0a}
\end{equation}%
\begin{equation}
\mathrm{sgn}\left( x\right) =\left\{ 
\begin{array}{lll}
1 & \text{if} & x>0 \\ 
0 & \text{if} & x=0 \\ 
-1 & \text{if} & x<0%
\end{array}%
\right. ,  \label{a4.1}
\end{equation}%
where $\sigma _{\mathrm{M}}$ is the world function of the $4$-dimensional
space-time of Minkowski. $\lambda _{0}$ is some elementary length. In such a
space-time geometry two connected equivalent timelike vectors $\mathbf{P}_{0}%
\mathbf{P}_{1}$ and $\mathbf{P}_{1}\mathbf{P}_{2}$ are described as follows 
\cite{R2007a} 
\begin{equation}
\mathbf{P}_{0}\mathbf{P}_{1}\mathrm{eqv}\mathbf{P}_{1}\mathbf{P}_{2}:\qquad 
\mathbf{P}_{0}\mathbf{P}_{1}=\left\{ \mu ,0,0,0\right\} ,\qquad \mathbf{P}%
_{1}\mathbf{P}_{2}=\left\{ \mu +\frac{3\lambda _{0}^{2}}{\mu },\lambda _{0}%
\sqrt{6+\frac{9\lambda _{0}^{2}}{\mu ^{2}}}\mathbf{n}\right\}  \label{a6.15}
\end{equation}%
where $\mathbf{n}$ is an arbitrary unit 3-vector. The quantity $\mu $ is the
length of the vector $\mathbf{P}_{0}\mathbf{P}_{1}$ (geometrical mass,
associated with the particle, which is described by the vector $\mathbf{P}%
_{0}\mathbf{P}_{1}$). We see that the spatial part of the vector $\mathbf{P}%
_{1}\mathbf{P}_{2}$ is determined to within the arbitrary 3-vector of the
length $\lambda _{0}\sqrt{6+\frac{9\lambda _{0}^{2}}{\mu ^{2}}}$. This
multivariance generates wobbling of the links of the world chain, consisting
of equivalent timelike vectors $...\mathbf{P}_{0}\mathbf{P}_{1}$, $\mathbf{P}%
_{1}\mathbf{P}_{2}$, $\mathbf{P}_{2}\mathbf{P}_{3}$,... Statistical
description of the chain with wobbling links coincides with the quantum
description of the particle with the mass $m=b\mu $, if the elementary
length $\lambda _{0}=\hbar ^{1/2}\left( 2bc\right) ^{-1/2}$, where $c$ is
the speed of the light, $\hbar $ is the quantum constant, and $b$ is some
universal constant, whose exact value is not determined \cite{R91}, because
the statistical description does not contain the quantity $b$. Thus, the
characteristic wobbling length is of the order of $\lambda _{0}$.

To explain the quantum description of the particle motion as a statistical
description of the multivariant classical motion, we should use the world
function (\ref{a4.1}). However, the form of the world function (\ref{a4.1})
is determined by the coincidence of the two descriptions only for the value $%
\sigma _{\mathrm{M}}>\sigma _{0}$, where the constant $\sigma _{0}$ is
determined via the mass $m_{\mathrm{L}}$ of the lightest massive particle
(electron) by means of the relation 
\begin{equation}
\sigma _{0}\leq \frac{\mu _{\mathrm{L}}^{2}}{2}-d=\frac{m_{\mathrm{L}}^{2}}{%
2b^{2}}-d=\frac{m_{\mathrm{L}}^{2}}{2b^{2}}-\frac{\hbar }{2bc}  \label{a4.2}
\end{equation}%
where $\mu _{\mathrm{L}}=m_{\mathrm{L}}/b$ is the geometrical mass of the
lightest massive particle (electron). The geometrical mass $\mu _{\mathrm{LM}%
}$ of the same particle, considered in the space-time geometry of Minkowski,
has the form%
\begin{equation*}
\mu _{\mathrm{LM}}=\sqrt{\mu _{\mathrm{L}}^{2}-2d}
\end{equation*}%
As far as $\sigma _{0}>0$, and, hence, $m_{\mathrm{L}}^{2}-b\hbar c^{-1}>0$,
we obtain the following estimation for the universal constant $b$%
\begin{equation}
b<\frac{m_{\mathrm{L}}^{2}c}{\hbar }\approx 2.4\times 10^{-17}\text{g/cm}.
\label{a4.3}
\end{equation}

Intensity of wobbling may be described by the multivariance vector $b_{%
\mathrm{m}}$, which is defined as follows. Let $\mathbf{P}_{1}\mathbf{P}_{2}$%
, $\mathbf{P}_{1}\mathbf{P}_{2}^{\prime }$ be two vectors which are
equivalent to the vector $\mathbf{P}_{0}\mathbf{P}_{2}$. Let 
\begin{equation*}
\mathbf{P}_{1}\mathbf{P}_{2}=\left\{ \mu +\frac{3\lambda _{0}^{2}}{\mu }%
,\lambda _{0}\sqrt{6+\frac{9\lambda _{0}^{2}}{\mu ^{2}}}\mathbf{n}\right\}
,\qquad \mathbf{P}_{1}\mathbf{P}_{2}^{\prime }=\left\{ \mu +\frac{3\lambda
_{0}^{2}}{\mu },\lambda _{0}\sqrt{6+\frac{9\lambda _{0}^{2}}{\mu ^{2}}}%
\mathbf{n}^{\prime }\right\}
\end{equation*}%
Let us consider the vector 
\begin{equation}
\mathbf{P}_{2}\mathbf{P}_{2}^{\prime }=\left\{ 0,\lambda _{0}\sqrt{6+\frac{%
9\lambda _{0}^{2}}{\mu ^{2}}}\left( \mathbf{n}^{\prime }-\mathbf{n}\right)
\right\}  \label{a4.3a}
\end{equation}%
which is a difference of vectors $\mathbf{P}_{1}\mathbf{P}_{2}$, $\mathbf{P}%
_{1}\mathbf{P}_{2}^{\prime }$. We consider the length $\left\vert \mathbf{P}%
_{2}\mathbf{P}_{2}^{\prime }\right\vert _{\mathrm{M}}$ of the vector $%
\mathbf{P}_{2}\mathbf{P}_{2}^{\prime }$ in the Minkowski space-time. We
obtain%
\begin{equation}
\left\vert \mathbf{P}_{2}\mathbf{P}_{2}^{\prime }\right\vert _{\mathrm{M}%
}^{2}=-\lambda _{0}^{2}\left( 6+\frac{9\lambda _{0}^{2}}{\mu ^{2}}\right)
\left( 2-2\mathbf{nn}^{\prime }\right)  \label{a4.3b}
\end{equation}%
The length of the vector (\ref{a4.3a}) is minimal at $\mathbf{n}=-\mathbf{n}%
^{\prime }$. At $\mathbf{n}=\mathbf{n}^{\prime }$ the length of the vector (%
\ref{a4.3a}) is maximal, and it is equal to zero. By definition the vector $%
\mathbf{P}_{2}\mathbf{P}_{2}^{\prime }$ at $\mathbf{n}=-\mathbf{n}^{\prime }$
is the multivariance 4-vector $b_{\mathrm{m}}$, which describes the
intensity of the multivariance. We have 
\begin{equation}
b_{\mathrm{m}}=\left\{ 0,\ 2\lambda _{0}\sqrt{6+\frac{9\lambda _{0}^{2}}{\mu
^{2}}}\mathbf{n}\right\} \qquad \left\vert b_{\mathrm{m}}\right\vert
^{2}=\left( b_{\mathrm{m}}.b_{\mathrm{m}}\right) =-4\lambda _{0}^{2}\left( 6+%
\frac{9\lambda _{0}^{2}}{\mu ^{2}}\right)  \label{a4.3c}
\end{equation}%
where $\mathbf{n}$ is an arbitrary unit 3-vector. The multivariance vector $%
b_{\mathrm{m}}$ is spacelike

In the case, when $\mu \gg \lambda _{0}$, the corresponding wobbling length 
\begin{equation*}
\lambda _{\mathrm{w}}=\frac{1}{2}\sqrt{\left\vert \left( b_{\mathrm{m}}.b_{%
\mathrm{m}}\right) \right\vert }\approx \sqrt{6}\lambda _{0}=\sqrt{6}\sqrt{%
\frac{\hbar }{2bc}}>\sqrt{3}\frac{\hbar }{m_{\mathrm{L}}c}=\sqrt{3}\lambda _{%
\mathrm{com}}
\end{equation*}%
where $\lambda _{\mathrm{com}}$ is the electron Compton wave length.

The relation (\ref{a4.3}) means that 
\begin{equation}
\sigma _{\mathrm{d}}=\sigma _{\mathrm{M}}+d,\ \ \ \ \text{if }\sigma _{%
\mathrm{M}}>\sigma _{0}  \label{a4.4}
\end{equation}%
For other values $\sigma _{\mathrm{M}}<\sigma _{0}$ the form of the world
function $\sigma _{\mathrm{d}}$ may distinguish from the relation (\ref{a4.4}%
). However, $\sigma _{\mathrm{d}}=0$, if $\sigma _{\mathrm{M}}=0$.

Two equivalent connected spacelike vectors $\mathbf{Q}_{0}\mathbf{Q}_{1}$, $%
\mathbf{Q}_{1}\mathbf{Q}_{2}$ have the form \cite{R2007a}%
\begin{equation}
\mathbf{Q}_{0}\mathbf{Q}_{1}=\left\{ 0,l,0,0\right\} ,\qquad \mathbf{Q}_{1}%
\mathbf{Q}_{2}=\left\{ \sqrt{\gamma _{2}^{2}+\gamma _{3}^{2}+6\lambda
_{0}^{2}+\frac{9\lambda _{0}^{4}}{l^{2}}},l,\gamma _{2},\gamma _{3}\right\}
\label{a6.50}
\end{equation}%
where constants $\gamma _{2}$ and $\gamma _{3}$ are arbitrary. The result is
obtained for the space-time geometry (\ref{a4.0}). Arbitrariness of
constants $\gamma _{2},\gamma _{3}$ generates multivariance of the vector $%
\mathbf{Q}_{1}\mathbf{Q}_{2}$ even in the space-time geometry of Minkowski,
where $\lambda _{0}=0$.

Vectors $\mathbf{Q}_{1}\mathbf{Q}_{2}$, $\mathbf{Q}_{1}\mathbf{Q}%
_{2}^{\prime }$ 
\begin{eqnarray*}
\mathbf{Q}_{1}\mathbf{Q}_{2} &=&\left\{ \sqrt{\gamma _{2}^{2}+\gamma
_{3}^{2}+6\lambda _{0}^{2}+\frac{9\lambda _{0}^{4}}{l^{2}}},l,\gamma
_{2},\gamma _{3}\right\} , \\
\mathbf{Q}_{1}\mathbf{Q}_{2}^{\prime } &=&\left\{ \sqrt{\gamma _{2}^{\prime
2}+\gamma _{3}^{\prime 2}+6\lambda _{0}^{2}+\frac{9\lambda _{0}^{4}}{l^{2}}}%
,l,\gamma _{2}^{\prime },\gamma _{3}^{\prime }\right\}
\end{eqnarray*}%
are equivalent to the vector $\mathbf{Q}_{0}\mathbf{Q}_{1}$. The difference $%
\mathbf{Q}_{2}\mathbf{Q}_{2}^{\prime }$ of two vectors $\mathbf{Q}_{1}%
\mathbf{Q}_{2}$, $\mathbf{Q}_{1}\mathbf{Q}_{2}^{\prime }$ has the form%
\begin{equation*}
\mathbf{Q}_{2}\mathbf{Q}_{2}^{\prime }=\left\{ \sqrt{\gamma _{2}^{\prime
2}+\gamma _{3}^{\prime 2}+6\lambda _{0}^{2}+\frac{9\lambda _{0}^{4}}{l^{2}}}-%
\sqrt{\gamma _{2}^{2}+\gamma _{3}^{2}+6\lambda _{0}^{2}+\frac{9\lambda
_{0}^{4}}{l^{2}}},0,\gamma _{2}^{\prime }-\gamma _{2},\gamma _{3}^{\prime
}-\gamma _{3}\right\}
\end{equation*}%
The vector $\mathbf{Q}_{2}\mathbf{Q}_{2}^{\prime }$ may be spacelike and
timelike. Its length has an extremum, if $\gamma _{2}^{\prime }=\gamma _{2}$
and $\gamma _{3}^{\prime }=\gamma _{3}$. In this case the length $\left\vert 
\mathbf{Q}_{2}\mathbf{Q}_{2}^{\prime }\right\vert ^{2}=0$

However, the length%
\begin{eqnarray*}
\left\vert \mathbf{Q}_{2}\mathbf{Q}_{2}^{\prime }\right\vert ^{2} &=&\left( 
\sqrt{\gamma _{2}^{\prime 2}+\gamma _{3}^{\prime 2}+6\lambda _{0}^{2}+\frac{%
9\lambda _{0}^{4}}{l^{2}}}-\sqrt{\gamma _{2}^{2}+\gamma _{3}^{2}+6\lambda
_{0}^{2}+\frac{9\lambda _{0}^{4}}{l^{2}}}\right) ^{2} \\
&&-\left( \gamma _{2}^{\prime }-\gamma _{2}\right) ^{2}-\left( \gamma
_{3}^{\prime }-\gamma _{3}\right) ^{2}
\end{eqnarray*}%
has neither maximum, nor minimum, and one cannot introduce the multivariance
vector of the type (\ref{a4.3c}). The multivariance of the spacelike vectors
equality is not introduced by the distortion $d$, defined by (\ref{a4.0a}).
It takes place already in the space-time of Minkowski. In the conventional
approach to the geometry of Minkowski one does not accept the multivariance
of spacelike vectors equivalence. Furthermore, the concept of multivariance
of two vectors parallelism (and equality) is absent at all in the
conventional approach to the geometry. For instance, when in the Riemannian
geometry the multivariance of parallelism of remote vectors appears, the
mathematicians prefer to deny at all the fernparallelism (parallelism of two
remote vectors), but not to introduce the concept of multivariance. This
circumstance is connected with the fact, that the multivariance may not
appear, if the geometry is constructed on the basis of a system of axioms .

The world chain, consisting of timelike equivalent vectors, imitates a world
line of a free particle. This fact seems to be rather reasonable.
Considering the vectors $\mathbf{Q}_{0}\mathbf{Q}_{1}$ and $\mathbf{Q}_{1}%
\mathbf{Q}_{2}$ in (\ref{a6.50}) from the viewpoint of the geometry of
Minkowski, we see that the vector $\mathbf{Q}_{1}\mathbf{Q}_{2}$ is obtained
from the vector $\mathbf{Q}_{0}\mathbf{Q}_{1}$ as a result of spatial
rotation and of an addition of some temporal component. One should expect,
that the world chain, consisting of spacelike equivalent vectors, imitates a
world line of a free particle, moving with a superluminal velocity. The
motion with the superluminal velocity seems to be unobservable. Such a
motion is considered to be impossible. However, if the spacelike world line
has a shape of a helix with timelike axis, such a situation may be
considered as a free rotating particle. The fact, that the particle rotates
with the superluminal velocity is not so important, if the helix axis is
timelike. The world line of a classical Dirac particle is a helix. It is not
very important, whether the rotation velocity is tardyon or not. Especially,
if we take into account that the Dirac equation describes the internal
degrees of freedom (rotation) nonrelativistically, (i.e. the description of
internal degrees of the classical Dirac particle is incorrect from the
viewpoint of the relativity theory).

We investigate now, whether a world chain of equivalent spacelike vectors
may form a helix with timelike axis. If it is possible, then we try to
investigate, under which world function such a situation is possible. We
consider the world function $\sigma _{\mathrm{d}}$ of the form 
\begin{equation}
\sigma _{\mathrm{d}}=\sigma _{\mathrm{M}}+d\cdot f\left( \frac{\sigma _{%
\mathrm{M}}}{\sigma _{0}}\right) ,\qquad f\left( x\right) =\mathrm{sgn}%
\left( x\right) ,\text{ \ \ \ if\ \ \ }\left\vert x\right\vert >1,\qquad
d=\lambda _{0}^{2}=\text{const}>0  \label{a4.5}
\end{equation}%
where the function $f\left( \frac{\sigma _{\mathrm{M}}}{\sigma _{0}}\right) $
should be determined from the condition, that the world chain, consisting of
spacelike links, forms a helix with timelike axis.

To estimate the form of $\sigma _{\mathrm{d}}$ as a function of $\sigma _{%
\mathrm{M}}$ at $\sigma _{\mathrm{M}}<\sigma _{0}$, we consider the chain,
consisting of equivalent spacelike vectors ... $\mathbf{P}_{0}\mathbf{P}_{1}$%
, $\mathbf{P}_{1}\mathbf{P}_{2}$, $\mathbf{P}_{2}\mathbf{P}_{3},$... We
suppose that the chain is a helix with timelike axis in the space-time. Let
the points $...P_{0},P_{1},..$. have the coordinates%
\begin{equation}
P_{k}=\left\{ kl_{0},R\cos \left( k\varphi -\varphi _{0}\right) ,R\sin
\left( k\varphi -\varphi _{0}\right) ,0\right\} ,\qquad k=...0,1,2,...
\label{a4.6}
\end{equation}%
All points (\ref{a4.6}) lie on a helix with timelike axis. In the space-time
of Minkowski the step of helix is equal to $2\pi l_{0}/\varphi $, and $R$ is
the radius of the helix. The constants $\varphi $ and $\varphi _{0}$ are
parameters of the helix. All vectors $\mathbf{P}_{k}\mathbf{P}_{k+1}$ have
the same length. Introducing designations \label{02beg}%
\begin{equation}
\phi =\frac{\varphi }{2},\qquad l_{1}=2R\sin \phi ,\qquad \varphi _{0}=\phi -%
\frac{\pi }{2}  \label{a4.7}
\end{equation}%
we obtain coordinates of vectors $\mathbf{P}_{k}\mathbf{P}_{k+1}$ in the
form 
\begin{equation}
\mathbf{P}_{k-1}\mathbf{P}_{k}=\left\{ l_{0},l_{1}\cos \left( 2k\phi \right)
,l_{1}\sin \left( 2k\phi \right) ,0\right\} ,\qquad k=...0,1,...
\label{a4.8}
\end{equation}%
where $l_{0},l_{1},\phi $ are parameters of the helix.

Let us investigate, under which conditions the relation $\mathbf{P}_{k-1}%
\mathbf{P}_{k}$eqv$\mathbf{P}_{k}\mathbf{P}_{k+1}$ takes place. We suppose
that all vectors of the helix are spacelike $\left\vert \mathbf{P}_{k}%
\mathbf{P}_{k+1}\right\vert ^{2}<0$. It is evident, that it is sufficient to
investigate the situation for the case $k=1$, when $\mathbf{P}_{0}\mathbf{P}%
_{1}$eqv$\mathbf{P}_{1}\mathbf{P}_{2}$. Let coordinates of vectors $\mathbf{P%
}_{0}\mathbf{P}_{1}$, $\mathbf{P}_{1}\mathbf{P}_{2}$ be 
\begin{equation}
\mathbf{P}_{0}\mathbf{P}_{1}=\left\{ l_{0},l_{1},0,0\right\} ,\qquad \mathbf{%
P}_{1}\mathbf{P}_{2}=\left\{ l_{0},l_{1}\cos \left( 2\phi \right) ,l_{1}\sin
\left( 2\phi \right) ,0\right\}  \label{a4.9}
\end{equation}%
In this case the coordinates of the points $P_{0},P_{1},P_{2}$ may be chosen
in the form%
\begin{equation}
P_{0}=\left\{ 0,0,0,0\right\} ,\qquad P_{1}=\left\{ l_{0},l_{1},0,0\right\}
,\qquad P_{2}=\left\{ 2l_{0},l_{1}\left( 1+\cos \left( 2\phi \right) \right)
,l_{1}\sin \left( 2\phi \right) ,0\right\} ,  \label{a4.11}
\end{equation}%
and the vector $\mathbf{P}_{0}\mathbf{P}_{2}$ has coordinates 
\begin{equation}
\mathbf{P}_{0}\mathbf{P}_{2}=\left\{ 2l_{0},l_{1}\left( 1+\cos \left( 2\phi
\right) \right) ,l_{1}\sin \left( 2\phi \right) ,0\right\}  \label{a4.12}
\end{equation}%
We choose the world function (\ref{a4.5}) in the form%
\begin{equation}
\sigma _{\mathrm{d}}=\sigma _{\mathrm{M}}+\lambda _{0}^{2}\left\{ 
\begin{array}{lll}
\mathrm{sgn}\left( \sigma _{\mathrm{M}}\right) & \text{if} & \left\vert
\sigma _{\mathrm{M}}\right\vert >\sigma _{0}>0 \\ 
\left( \frac{\sigma _{\mathrm{M}}}{\sigma _{0}}\right) ^{3} & \text{if} & 
\left\vert \text{\ }\sigma _{\mathrm{M}}\right\vert <\sigma _{0}%
\end{array}%
\right.  \label{a4.14}
\end{equation}%
and introduce the quantity 
\begin{equation}
\varkappa =\frac{\sigma _{0}}{\lambda _{0}^{2}}  \label{a4.14a}
\end{equation}

Thus, we have 
\begin{equation}
\sigma _{\mathrm{d}}=\sigma _{\mathrm{M}}+d\left( \sigma _{\mathrm{M}%
}\right) ,\qquad d\left( \sigma _{\mathrm{M}}\right) =\lambda
_{0}^{2}f\left( \frac{\sigma _{\mathrm{M}}}{\sigma _{0}}\right) ,\qquad
f\left( x\right) =\left\{ 
\begin{array}{lll}
1 & \text{if} & x\geq 1 \\ 
x^{3} & \text{if} & -1<x<1 \\ 
-1 & \text{if} & x\leq -1%
\end{array}%
\right.  \label{a4.15}
\end{equation}%
The space-time geometry (\ref{a4.15}) is a special case of the space-time
geometry (\ref{a4.4}). We do not pretend to the claim, that (\ref{a4.15}) is
the world function of true space-time geometry. We shall show only that in
the space-time geometry (\ref{a4.15}) the spacelike vectors (\ref{a4.9}) may
be equivalent at some proper choice of parameters $l_{0}$, $l_{1}$ and $%
\varphi $.

In our calculations we shall use two geometries: the geometry $\mathcal{G}_{%
\mathrm{M}}$ of Minkowski and the space-time geometry $\mathcal{G}_{\mathrm{d%
}}$, described by the world function $\sigma _{\mathrm{d}}$, determined by (%
\ref{a4.15}). Then expressions of the geometry $\mathcal{G}_{\mathrm{d}}$
may be reduced to expressions of the geometry $\mathcal{G}_{\mathrm{M}}$ by
means of relations%
\begin{equation}
\left\vert \mathbf{P}_{0}\mathbf{P}_{1}\right\vert ^{2}=\left\vert \mathbf{P}%
_{0}\mathbf{P}_{1}\right\vert _{\mathrm{M}}^{2}+2d\left( \sigma _{\mathrm{M}%
}\left( P_{0},P_{1}\right) \right)  \label{a4.15a}
\end{equation}%
\begin{equation}
\left( \mathbf{P}_{0}\mathbf{P}_{1}.\mathbf{Q}_{0}\mathbf{Q}_{1}\right)
=\left( \mathbf{P}_{0}\mathbf{P}_{1}.\mathbf{Q}_{0}\mathbf{Q}_{1}\right) _{%
\mathrm{M}}+w\left( P_{0},P_{1},Q_{0},Q_{1}\right)  \label{a4.15b}
\end{equation}%
\begin{eqnarray}
&&w\left( P_{0},P_{1},Q_{0},Q_{1}\right) =d\left( \sigma _{\mathrm{M}}\left(
P_{0},Q_{1}\right) \right) +d\left( \sigma _{\mathrm{M}}\left(
P_{1},Q_{0}\right) \right)  \notag \\
&&-d\left( \sigma _{\mathrm{M}}\left( P_{0},Q_{0}\right) \right) -d\left(
\sigma _{\mathrm{M}}\left( P_{1},Q_{1}\right) \right)  \label{a4.15c}
\end{eqnarray}%
The geometrical relations in $\mathcal{G}_{\mathrm{d}}$ are expressed via
the same relations, written in $\mathcal{G}_{\mathrm{M}}$ with additional
terms, containing the distortion $d$. This additional terms in dynamic
relations are interpreted as additional metric interactions, acting on a
particle, when the real space-time geometry $\mathcal{G}_{\mathrm{d}}$ is
considered to be the geometry $\mathcal{G}_{\mathrm{M}}$. Appearance of
additional interactions reminds appearance of inertial forces at a use of
accelerated coordinate systems instead of inertial ones.

Condition $\mathbf{P}_{0}\mathbf{P}_{1}$eqv$\mathbf{P}_{1}\mathbf{P}_{2}$ of
equivalence of vectors $\mathbf{P}_{0}\mathbf{P}_{1}$, $\mathbf{P}_{1}%
\mathbf{P}_{2}$ is written in the form of two equations%
\begin{equation}
\left( \mathbf{P}_{0}\mathbf{P}_{1}.\mathbf{P}_{1}\mathbf{P}_{2}\right)
=\left( \mathbf{P}_{0}\mathbf{P}_{1}.\mathbf{P}_{1}\mathbf{P}_{2}\right) _{%
\mathrm{M}}+w\left( P_{0},P_{1},P_{1},P_{2}\right) =\left\vert \mathbf{P}_{0}%
\mathbf{P}_{1}\right\vert _{\mathrm{M}}^{2}+2d\left( \sigma _{\mathrm{M}%
}\left( P_{0},P_{1}\right) \right)  \label{a4.16}
\end{equation}%
\begin{equation}
\left\vert \mathbf{P}_{0}\mathbf{P}_{1}\right\vert _{\mathrm{M}%
}^{2}+2d\left( \sigma _{\mathrm{M}}\left( P_{0},P_{1}\right) \right)
=\left\vert \mathbf{P}_{1}\mathbf{P}_{2}\right\vert _{\mathrm{M}%
}^{2}+2d\left( \sigma _{\mathrm{M}}\left( P_{1},P_{2}\right) \right)
\label{a4.17}
\end{equation}%
where index 'M' means, that the corresponding quantities are calculated in $%
\mathcal{G}_{\mathrm{M}}$. The function $d$ is determined by the relation (%
\ref{a4.15}), and the quantity $w$ is determined by the relation 
\begin{equation}
w\left( P_{0},P_{1},P_{1},P_{2}\right) =d\left( \sigma _{\mathrm{M}}\left(
P_{0},P_{2}\right) \right) -d\left( \sigma _{\mathrm{M}}\left(
P_{0},P_{1}\right) \right) -d\left( \sigma _{\mathrm{M}}\left(
P_{1},P_{2}\right) \right)  \label{a4.18}
\end{equation}%
which follows from the definition of the scalar product (\ref{a4.15c}).
Using the conventional relations for the scalar product in $\mathcal{G}_{%
\mathrm{M}}$, we can rewrite the relations (\ref{a4.16}), (\ref{a4.17}) in
the form 
\begin{equation}
l_{0}^{2}-l_{1}^{2}\cos \left( 2\phi \right) +w\left(
P_{0},P_{1},P_{1},P_{2}\right) =l_{0}^{2}-l_{1}^{2}+2d\left( \sigma _{%
\mathrm{M}}\left( P_{0},P_{1}\right) \right)  \label{a4.19}
\end{equation}%
\begin{equation}
l_{0}^{2}-l_{1}^{2}=l_{0}^{2}-l_{1}^{2}\left( \cos ^{2}\left( 2\phi \right)
+\sin ^{2}\left( 2\phi \right) \right)  \label{a4.20}
\end{equation}%
where%
\begin{equation}
w\left( P_{0},P_{1},P_{1},P_{2}\right) =d\left( 2l_{1}^{2}\sin ^{2}\phi
+2\left( l_{0}^{2}-l_{1}^{2}\right) \right) -2d\left( \frac{%
l_{0}^{2}-l_{1}^{2}}{2}\right)  \label{a4.21}
\end{equation}

To obtain the relation (\ref{a4.21}) from (\ref{a4.18}), we use the relations%
\begin{equation}
\left\vert \mathbf{P}_{0}\mathbf{P}_{1}\right\vert _{\mathrm{M}%
}^{2}=\left\vert \mathbf{P}_{1}\mathbf{P}_{2}\right\vert _{\mathrm{M}%
}^{2}=2\sigma _{\mathrm{M}}\left( P_{0},P_{1}\right)
=l_{0}^{2}-l_{1}^{2}\equiv l^{2}  \label{a4.22}
\end{equation}%
\begin{equation}
\frac{1}{2}\left\vert \mathbf{P}_{0}\mathbf{P}_{2}\right\vert _{\mathrm{M}%
}^{2}=\sigma _{\mathrm{M}}\left( P_{0},P_{2}\right)
=2l_{0}^{2}-l_{1}^{2}\left( 1+\cos \left( 2\phi \right) \right)
=2l_{1}^{2}\sin ^{2}\phi +2l^{2}  \label{a4.23}
\end{equation}%
The equation (\ref{a4.20}) is the identity.

Let us introduce pure quantities $\nu $, $a$, defining them by relations%
\begin{equation}
l^{2}=l_{0}^{2}-l_{1}^{2}=-2\nu \sigma _{0},\qquad \nu >0  \label{a4.24}
\end{equation}%
\begin{equation}
a=\frac{2l_{1}^{2}}{\sigma _{0}}\sin ^{2}\phi ,\qquad \varkappa =\frac{%
\sigma _{0}}{\lambda _{0}^{2}}  \label{a4.25}
\end{equation}%
Then the equation (\ref{a4.19}) takes the form%
\begin{equation}
\varkappa a+f\left( a-4\nu \right) =-4f\left( \nu \right)  \label{a4.25a}
\end{equation}%
where the function $f$ is defined by the relation (\ref{a4.15})%
\begin{equation}
f\left( \nu \right) =\frac{1}{\lambda _{0}^{2}}d\left( \sigma _{0}\nu
\right) =\left\{ 
\begin{array}{lll}
\mathrm{sgn}\left( \nu \right) & \text{if} & \left\vert \nu \right\vert >1
\\ 
\nu ^{3} & \text{if} & \left\vert \nu \right\vert <1%
\end{array}%
\right.  \label{a4.26}
\end{equation}%
and the constant $\varkappa $ is defined by the relation (\ref{a4.14a}).

Let us note, that in the case, when $f\left( \nu \right) $ is a linear
function $f\left( \nu \right) =\nu $, for $\nu \in \left[ -1,1\right] $, the
equation (\ref{a4.25a}) has the unique solution $a=0$. The solution with $a=%
\frac{2l_{1}^{2}}{\sigma _{0}}\sin ^{2}\phi =0$ describes a straight but not
a helix.

Considering solutions of equation (\ref{a4.25a}) with respect to $a=a\left(
\nu \right) $, we are interested in positive values of $a$, because the
quantity $a$ is nonnegative by definition (\ref{a4.25}). At $\varkappa =1$
numerical solutions of equation (\ref{a4.25a}) with respect to $a$ are
presented in the form%
\begin{equation*}
\begin{array}{ll}
\nu & a\left( \nu \right) \\ 
0 & 0 \\ 
0.1 & 0.04191 \\ 
0.2 & 0.19236 \\ 
0.3 & 0.40137%
\end{array}%
\qquad 
\begin{array}{ll}
\nu & a\left( \nu \right) \\ 
0.4 & 0.63701 \\ 
0.5 & 0.5 \\ 
0.6 & 0.136 \\ 
0.63 & 0%
\end{array}%
\qquad 
\begin{array}{ll}
\nu & a\left( \nu \right) \\ 
-0.63 & 0 \\ 
-0.7 & 0.372 \\ 
-0.8 & 1.048 \\ 
-0.9 & 1.\,\allowbreak 916%
\end{array}%
\qquad 
\begin{array}{ll}
\nu & a\left( \nu \right) \\ 
-0.956\,47 & 2 \\ 
-0.974\,35 & 2.7 \\ 
-0.991\,60 & 2.9 \\ 
-1 & 3%
\end{array}%
\end{equation*}

According to (\ref{a4.7}), (\ref{a4.24}) and (\ref{a4.25}) we have the
following relations for the helix radius $R$ 
\begin{equation}
\sin \phi =\sqrt{\frac{a\sigma _{0}}{2l_{1}^{2}}},\qquad R=\frac{l_{1}}{%
2\sin \phi }=\frac{l_{1}^{2}}{\sqrt{2a\sigma _{0}}}  \label{a4.27}
\end{equation}%
We obtain the helix step $S$ in the form%
\begin{equation}
S=\frac{\pi }{\phi }l_{0}=\frac{\pi l_{0}}{\arcsin \sqrt{\frac{a\sigma _{0}}{%
2l_{1}^{2}}}}=\frac{\pi \sqrt{l_{1}^{2}-2\sigma _{0}\nu }}{\arcsin \sqrt{%
\frac{a\sigma _{0}}{2l_{1}^{2}}}}  \label{a4.28}
\end{equation}

Negative values of $\nu $ correspond to helix with timelike vectors $\mathbf{%
P}_{k-1}\mathbf{P}_{k}$. Positive solutions of equation (\ref{a4.25a}) take
place only for $\nu \in \left( 0,0.63\right) $ (spacelike vectors) and $\nu
\in \left( -0.63,-1\right) $ (timelike vectors). The values of parameter $\
a $ belong to intervals

\begin{equation}
a\in \left[ 0,0.695\right] ,\qquad a\in \left( 0,3\right)  \label{a4.38}
\end{equation}%
for spacelike and timelike vectors correspondingly.

Thus, we see that in the space-time geometry with the world function (\ref%
{a4.15}) the spatial evolution, determined by the spacelike vectors $\mathbf{%
P}_{k}\mathbf{P}_{k+1}$, may lead to a helical world chain with timelike
axis. However, equivalence of spacelike vectors $\mathbf{P}_{k}\mathbf{P}%
_{k+1}$ is multivariant even in the space-time of Minkowski. It is valid for
the space-time geometry (\ref{a4.15}) also. As a result the wobbling of the
spacelike vectors appears. It may lead to destruction of the helix.

In reality the conditions $\mathbf{P}_{0}\mathbf{P}_{1}\mathrm{eqv}\mathbf{P}%
_{1}\mathbf{P}_{2}$ determines vector $\mathbf{P}_{1}\mathbf{P}_{2}\mathbf{\ 
}$to within the vector $\alpha =\left\{ \alpha _{0},\mathbf{\alpha }\right\} 
$, and we have instead of equations (\ref{a4.9}) 
\begin{equation}
\mathbf{P}_{0}\mathbf{P}_{1}=l,\qquad \mathbf{P}_{1}\mathbf{P}_{2}=q+\alpha
\label{a4.39}
\end{equation}%
where $l$, $q$, $\alpha $ are 4-vectors with coordinates%
\begin{equation}
l=\left\{ l_{0},l_{1},0,0\right\} ,\qquad q=\left\{ l_{0},l_{1}\cos \left(
2\phi \right) ,l_{1}\sin \left( 2\phi \right) ,0\right\} ,\qquad \alpha
=\left\{ \alpha _{0},\alpha _{1},\alpha _{2},\alpha _{3}\right\}
\label{a4.40}
\end{equation}

Instead of equations (\ref{a4.19}) -- (\ref{a4.21}) we have the following
equations%
\begin{equation}
\alpha ^{2}+2\left( q.\alpha \right) =0  \label{a4.41}
\end{equation}%
\begin{equation}
2l_{1}^{2}\sin ^{2}\phi +\left( l.\alpha \right) +\lambda _{0}^{2}f\left( 
\frac{2l^{2}+2l_{1}^{2}\sin ^{2}\phi +\left( l.\alpha \right) }{\sigma _{0}}%
\right) -4\lambda _{0}^{2}f\left( \frac{l^{2}}{2\sigma _{0}}\right) =0
\label{a4.42}
\end{equation}%
where $\left( l.\alpha \right) $ and $\left( q.\alpha \right) $ mean scalar
products of vectors $l,q,\alpha $ in the space-time of Minkowski. The
relation (\ref{a4.25a}) is the necessary condition of the fact, that $\alpha
=0$ is a solution of equations (\ref{a4.41}), (\ref{a4.42}). We obtain from (%
\ref{a4.41}) 
\begin{equation}
\alpha _{0}=-q_{0}\pm \sqrt{q_{0}^{2}+\mathbf{\alpha }^{2}+2\mathbf{q\alpha }%
}=-l_{0}\pm \sqrt{l_{0}^{2}+\mathbf{\alpha }^{2}+2\mathbf{q\alpha }}
\label{a4.43}
\end{equation}%
where $\mathbf{q\alpha }$ means the scalar product of 3-vectors $\mathbf{q}$
and $\mathbf{\alpha }$.

Taking into account the relation (\ref{a4.25a}), we obtain from relation (%
\ref{a4.42}) 
\begin{equation}
\left( l.\alpha \right) +\lambda _{0}^{2}\left( f\left( \frac{%
2l^{2}+2l_{1}^{2}\sin ^{2}\phi +\left( l.\alpha \right) }{\sigma _{0}}%
\right) -f\left( \frac{2l^{2}+2l_{1}^{2}\sin ^{2}\phi }{\sigma _{0}}\right)
\right) =0  \label{a4.44}
\end{equation}%
Supposing, that $\left( l.\alpha \right) $ is a small quantity we obtain
from (\ref{a4.44}) by means of (\ref{a4.43})%
\begin{equation}
\left( l_{0}\left( -l_{0}\pm \sqrt{l_{0}^{2}+\mathbf{\alpha }^{2}+2\mathbf{%
q\alpha }}\right) -\mathbf{l\alpha }\right) \left( 1+\frac{\lambda _{0}^{2}}{%
\sigma _{0}}f^{\prime }\left( \frac{2l^{2}+2l_{1}^{2}\sin ^{2}\phi }{\sigma
_{0}}\right) \right) =0  \label{a4.45}
\end{equation}

The relation (\ref{a4.45}) may be transformed to the equation%
\begin{equation}
\left( 1-\frac{\mathbf{l}^{2}}{l_{0}^{2}}\right) \left( \mathbf{\alpha }%
_{\parallel }+\frac{\mathbf{q}_{\parallel }-2\mathbf{l}}{1-\frac{\mathbf{l}%
^{2}}{l_{0}^{2}}}\right) ^{2}\mathbf{+}\left( \mathbf{\alpha }_{\perp }+%
\mathbf{q}_{\perp }\right) ^{2}=\frac{\left( \mathbf{q}_{\parallel }-2%
\mathbf{l}\right) ^{2}}{\left( 1-\frac{\mathbf{l}^{2}}{l_{0}^{2}}\right) }+%
\mathbf{q}_{\perp }^{2}  \label{a4.46}
\end{equation}%
where%
\begin{equation}
\mathbf{\alpha }=\mathbf{\alpha }_{\parallel }\mathbf{+\alpha }_{\perp
},\qquad \mathbf{\alpha }_{\parallel }=\frac{\mathbf{l}\left( \mathbf{%
l\alpha }\right) }{\mathbf{l}^{2}},\qquad \mathbf{q}_{\parallel }=\frac{%
\mathbf{l}\left( \mathbf{lq}\right) }{\mathbf{l}^{2}},\qquad \mathbf{q}%
_{\perp }=\mathbf{q}-\mathbf{q}_{\parallel }  \label{a4.47}
\end{equation}%
As far as $\mathbf{l}^{2}>l_{0}^{2}$, we obtain, that $1-\mathbf{l}%
^{2}/l_{0}^{2}<0$, and the surface (\ref{a4.46}) is a hyperboloid in the
3-space of quantities $\alpha _{1},\alpha _{2},\alpha _{3}$. It means that
solutions of the equations (\ref{a4.43}), (\ref{a4.44}) may deflect
arbitrarily far from the helix solution (\ref{a4.9}). This deflection is a
manifestation of the multivariance of the space-time geometry.

\section{Stabilization of the spacelike world chain}

Suppression of multivariance and stabilization of the world chain,
consisting of spacelike vectors, can be achieved, if we consider the world
chain with composed links, whose skeleton consists of three points $\left\{
P_{k},P_{k+1},Q_{k+1}\right\} $, $k=...1.2,...$ (see figure 1). Let $\mathbf{%
P}_{k}\mathbf{P}_{k+1}$ be a spacelike vector, whereas the vector $\mathbf{P}%
_{k}\mathbf{Q}_{k+1}$ be a timelike vector in $\mathcal{G}_{\mathrm{M}}$. To
investigate the effect of stabilization, it is sufficient to consider the
points $P_{0},P_{1},P_{2},Q_{1},Q_{2}$, having coordinates%
\begin{eqnarray}
P_{0} &=&\left\{ 0\right\} ,\qquad P_{1}=\left\{ l\right\} ,\qquad
P_{2}=\left\{ l\mathbf{+}q+\alpha \right\} ,  \notag \\
Q_{1} &=&\left\{ s\right\} ,\qquad Q_{2}=\left\{ s+\rho +l+\beta \right\}
\label{a5.0}
\end{eqnarray}%
The following vectors are associated with these points of the skeletons%
\begin{eqnarray}
\mathbf{P}_{0}\mathbf{P}_{1} &=&l,\qquad \mathbf{P}_{1}\mathbf{P}%
_{2}=q+\alpha ,\qquad \mathbf{P}_{0}\mathbf{P}_{2}=l\mathbf{+}q+\alpha ,
\label{a5.1} \\
\mathbf{P}_{0}\mathbf{Q}_{1} &=&s,\qquad \mathbf{P}_{1}\mathbf{Q}_{2}=s+\rho
+\beta ,\qquad \mathbf{P}_{0}\mathbf{Q}_{2}=s+\rho +l+\beta ,  \label{a5.2}
\\
\mathbf{P}_{1}\mathbf{Q}_{1} &=&s-l,\qquad \mathbf{P}_{2}\mathbf{Q}%
_{2}=s+\rho -q+\gamma ,\qquad \mathbf{Q}_{1}\mathbf{Q}_{2}=\rho +l+\beta ,
\label{a5.3} \\
\mathbf{Q}_{1}\mathbf{P}_{2} &=&l+q-s+\alpha ,\qquad \gamma =\beta -\alpha 
\notag
\end{eqnarray}%
where the quantities $l,q,s,\rho $ are the given 4-vectors, whereas the
quantities $\alpha ,\beta ,\gamma =\beta -\alpha $ are 4-vectors, which are
to be determined from the condition 
\begin{equation}
\left\{ P_{0},P_{1},Q_{1}\right\} \mathrm{eqv}\left\{
P_{1},P_{2},Q_{2}\right\}  \label{a5.3a}
\end{equation}

Expressions for 4-vectors $q$ and $\rho $ are chosen in such a way, that
vectors $\mathbf{P}_{1}\mathbf{P}_{2}$ and $\mathbf{P}_{1}\mathbf{Q}_{2}$
(at $\alpha =\beta =0$) were a result of rotation of vectors $\mathbf{P}_{0}%
\mathbf{P}_{1}$ and $\mathbf{P}_{0}\mathbf{Q}_{1}$ in the plane $x^{1}x^{2}$
by the angle $2\phi $. The quantities 
\begin{eqnarray}
s &=&\left\{ s_{0},s_{\perp }\cos \phi ,s_{\perp }\sin \phi ,s_{3}\right\}
\qquad q=\left\{ l_{0},l_{1}\cos \left( 2\phi \right) ,l_{1}\sin \left(
2\phi \right) ,0\right\}   \label{b5.4a} \\
\rho  &=&\left\{ 0,-2s_{\perp }\sin \phi \sin \left( 2\phi \right)
,2s_{\perp }\sin \phi \cos \left( 2\phi \right) ,0\right\} ,\qquad l=\left\{
l_{0},l_{1},0,0\right\}   \label{b5.4b}
\end{eqnarray}%
are supposed to be given. The 4-vectors 
\begin{equation}
\alpha =\left\{ \alpha _{0},\alpha _{1},\alpha _{2},\alpha _{3}\right\}
=\left\{ \alpha _{0},\mathbf{\alpha }\right\} ,\qquad \beta =\left\{ \beta
_{0},\beta _{1},\beta _{2},\beta _{3}\right\} =\left\{ \beta _{0},\mathbf{%
\beta }\right\}   \label{b5.5}
\end{equation}%
are to be determined from the relations (\ref{a5.3a}).

The 4-vectors $l$ and $q$ coincide with vectors (\ref{a4.9}). We are
interested in the following question, whether the stabilizing vector $%
\mathbf{P}_{0}\mathbf{Q}_{1}=s$ can be chosen in such a way, that equations (%
\ref{a5.3a}) have the unique solution $\alpha =\beta =0$. If such a
stabilizing vector $\frac{{}}{{}}$exists, the world chain will have a shape
of a helix without wobbling. It may be, that the complete stabilization is
impossible. Then, maybe, a partial stabilization is possible, when the
quantities $\alpha $, $\beta $ are small, although they do not vanish. In
any case the problem of the stabilizing vector existence is a pure
mathematical problem.

Solving this problem, we shall use relations (\ref{a4.15a}), (\ref{a4.15b})
to reduce all geometrical relations to the geometrical relations in the
space-time of Minkowski. Working in the space-time of Minkowski, we shall
use the conventional covariant formalism, where the expressions of the type $%
\alpha ^{2}$ and $\left( \alpha .\beta \right) $ mean%
\begin{eqnarray}
\alpha ^{2} &\equiv &\alpha _{0}^{2}-\mathbf{\alpha }^{2}\equiv \alpha
_{0}^{2}-\alpha _{1}^{2}-\alpha _{2}^{2}-\alpha _{3}^{2},  \label{b5.6} \\
\left( \alpha .\beta \right) &\mathbf{\equiv }&\alpha _{0}\beta _{0}-\mathbf{%
\alpha \beta \equiv }\alpha _{0}\beta _{0}-\alpha _{1}\beta _{1}-\alpha
_{2}\beta _{2}-\alpha _{3}\beta _{3}  \label{b5.7}
\end{eqnarray}%
Index "\textrm{M" }will be omitted for brevity.

It follows from the condition $\mathbf{P}_{0}\mathbf{P}_{1}$eqv$\mathbf{P}%
_{1}\mathbf{P}_{2}$%
\begin{eqnarray}
l^{2} &=&\left( q+\alpha \right) ^{2}  \label{b5.8} \\
\left( l.q+\alpha \right) +w\left( P_{0},P_{1},P_{1},P_{2}\right)
&=&l^{2}+2d\left( \frac{s^{2}}{2}\right)  \label{b5.9}
\end{eqnarray}%
where%
\begin{eqnarray}
&&w\left( P_{0},P_{1},P_{1},P_{2}\right)  \notag \\
&=&d\left( \sigma _{\mathrm{M}}\left( P_{0},P_{2}\right) \right) +d\left(
\sigma _{\mathrm{M}}\left( P_{1},P_{1}\right) \right) -d\left( \sigma _{%
\mathrm{M}}\left( P_{0},P_{1}\right) \right) -d\left( \sigma _{\mathrm{M}%
}\left( P_{1},P_{2}\right) \right)  \notag \\
&=&\lambda _{0}^{2}f\left( \frac{\left( l\mathbf{+}q+\alpha \right) ^{2}}{%
2\sigma _{0}}\right) -2\lambda _{0}^{2}f\left( \frac{l^{2}}{2\sigma _{0}}%
\right)  \label{b5.10}
\end{eqnarray}%
After transformations we obtain%
\begin{equation}
\alpha ^{2}+2\left( q.\alpha \right) =0  \label{b5.11}
\end{equation}%
\begin{equation}
2l_{1}^{2}\sin ^{2}\phi +\left( l.\alpha \right) +\lambda _{0}^{2}f\left( 
\frac{2l^{2}+2l_{1}^{2}\sin ^{2}\phi +\left( l.\alpha \right) }{\sigma _{0}}%
\right) -4\lambda _{0}^{2}f\left( \frac{l^{2}}{2\sigma _{0}}\right) =0
\label{b5.12}
\end{equation}%
These equations coincide with (\ref{a4.41}), (\ref{a4.42}). If $\alpha =0$
the equations (\ref{b5.11}), (\ref{b5.12}) coincide with (\ref{a4.20}), (\ref%
{a4.25a}) respectively.

We obtain from the condition $\mathbf{P}_{0}\mathbf{Q}_{1}$eqv$\mathbf{P}_{1}%
\mathbf{Q}_{2}$%
\begin{equation}
s^{2}=\left( s+\rho +\beta \right) ^{2}  \label{b5.14}
\end{equation}%
\begin{equation}
\left( s+\rho +\beta .s\right) +w\left( P_{0},Q_{1},P_{1},Q_{2}\right)
=s^{2}+2d\left( \frac{s^{2}}{2}\right)  \label{b5.15}
\end{equation}%
where%
\begin{eqnarray*}
&&w\left( P_{0},Q_{1},P_{1},Q_{2}\right) \\
&=&d\left( \sigma _{\mathrm{M}}\left( P_{0},Q_{2}\right) \right) +d\left(
\sigma _{\mathrm{M}}\left( Q_{1},P_{1}\right) \right) -d\left( \sigma _{%
\mathrm{M}}\left( P_{0},P_{1}\right) \right) -d\left( \sigma _{\mathrm{M}%
}\left( Q_{1},Q_{2}\right) \right) \\
&=&d\left( \frac{\left( s+\rho +l+\beta \right) ^{2}}{2}\right) +d\left( 
\frac{\left( s-l\right) ^{2}}{2}\right) -d\left( \frac{l^{2}}{2}\right)
-d\left( \frac{\left( \rho +l+\beta \right) ^{2}}{2}\right)
\end{eqnarray*}

The equations (\ref{b5.14}) and (\ref{b5.15}) are transformed to the form%
\begin{equation}
\rho ^{2}+2\left( \rho .s\right) +2\left( s+\rho .\beta \right) +\beta ^{2}=0
\label{b5.16}
\end{equation}%
\begin{equation*}
\left( \rho +\beta .s\right) +d\left( \frac{\left( s+\rho +l+\beta \right)
^{2}}{2}\right) +d\left( \frac{\left( s-l\right) ^{2}}{2}\right) -d\left( 
\frac{l^{2}}{2}\right)
\end{equation*}%
\begin{equation}
-d\left( \frac{\left( \rho +l+\beta \right) ^{2}}{2}\right) -2d\left( \frac{%
s^{2}}{2}\right) =0  \label{b5.17}
\end{equation}

Let us suppose that the stabilizing vector $s$ is long in the sense that 
\begin{equation}
s^{2}\gg \sigma _{0}  \label{b5.17a}
\end{equation}%
Then in (\ref{b5.17}) the functions $d$, which contains $s$ in its argument
will be equal to $\lambda _{0}^{2}$ and all terms, containing $s$ compensate
each other. The necessary condition of the fact, that $\beta =0$ is a
solution of equations (\ref{b5.16}), (\ref{b5.17}), has the form%
\begin{equation}
\rho ^{2}+2\left( \rho .s\right) =0  \label{b5.18}
\end{equation}%
\begin{equation}
\left( \rho .s\right) -d\left( \frac{l^{2}}{2}\right) -d\left( \frac{\left(
\rho +l\right) ^{2}}{2}\right) =0  \label{b5.19}
\end{equation}%
The equation (\ref{b5.18}) is satisfied identically by the choice (\ref%
{b5.4a}), (\ref{b5.4b}) of vectors $s$ and $\rho $.

We obtain from the condition $\mathbf{P}_{1}\mathbf{Q}_{1}$eqv$\mathbf{P}_{2}%
\mathbf{Q}_{2}$%
\begin{equation}
\left( s-l\right) ^{2}=\left( s+\rho -q+\gamma \right) ^{2}  \label{b5.23}
\end{equation}%
\begin{equation}
\left( s-l.s+\rho -q+\gamma \right) +w\left( P_{1},Q_{1},P_{2},Q_{2}\right)
=\left( s-l\right) ^{2}+2d\left( \frac{\left( s-l\right) ^{2}}{2}\right)
\label{b5.24}
\end{equation}%
\begin{equation*}
w\left( P_{1},Q_{1},P_{2},Q_{2}\right) =d\left( \sigma _{\mathrm{M}}\left(
P_{1},Q_{2}\right) \right) +d\left( \sigma _{\mathrm{M}}\left(
Q_{1},P_{2}\right) \right) -d\left( \sigma _{\mathrm{M}}\left(
P_{1},P_{2}\right) \right) -d\left( \sigma _{\mathrm{M}}\left(
Q_{1},Q_{2}\right) \right)
\end{equation*}%
\begin{equation}
=d\left( \frac{\left( s+\rho +\beta \right) ^{2}}{2}\right) +d\left( \frac{%
\left( l+q-s+\alpha \right) ^{2}}{2}\right) -d\left( \frac{l^{2}}{2}\right)
-d\left( \frac{\left( \rho +l+\beta \right) ^{2}}{2}\right)  \label{b5.26}
\end{equation}

The necessary conditions of the fact, that $\gamma =\beta -\alpha =0$ is a
solution of equations (\ref{b5.23}), (\ref{b5.24}), have the form%
\begin{equation}
\left( s-l\right) ^{2}=\left( s+\rho -q\right) ^{2}  \label{b5.27}
\end{equation}%
\begin{equation}
\left( s-l.s+\rho -q\right) -d\left( \frac{l^{2}}{2}\right) -d\left( \frac{%
\left( \rho +l\right) ^{2}}{2}\right) =0  \label{b5.28}
\end{equation}

The equation (\ref{b5.27}) is satisfied identically by the relations (\ref%
{b5.4a}), (\ref{b5.4b}). The difference of equations (\ref{b5.19}) and (\ref%
{b5.28}) leads to the equation

\begin{equation}
\left( \rho .s\right) =\left( s-l.s+\rho -q\right)  \label{b5.29}
\end{equation}%
Let us substitute expressions for $\rho ,s,l,q$ , determined by the
relations (\ref{b5.4a}), (\ref{b5.4b}), in (\ref{b5.29}). After
transformations we obtain the connection between the quantities $s_{\perp
},l_{1}$ and $\phi $ in the form%
\begin{equation}
s_{\perp }=l_{1}\frac{1-2\sin ^{2}\phi }{\left( 1-4\sin ^{2}\phi \right)
\cos \phi }  \label{b5.30}
\end{equation}

The equation (\ref{b5.19}) by means of (\ref{b5.30}) is reduced to the form%
\begin{eqnarray}
&&\frac{2l_{1}^{2}\left( 1-2\sin ^{2}\phi \right) ^{2}}{\left( 1-3\sin
^{2}\phi \right) ^{2}\left( 1-\sin ^{2}\phi \right) }\sin ^{2}\phi -\lambda
_{0}^{2}f\left( \frac{l^{2}}{2\sigma _{0}}\right)  \notag \\
&&-\lambda _{0}^{2}f\left( \frac{l^{2}+4l_{1}^{2}\frac{\left( 1-2\sin
^{2}\phi \right) }{\left( 1-3\sin ^{2}\phi \right) }\sin ^{2}\phi
-16l_{1}^{2}\frac{\left( 1-2\sin ^{2}\phi \right) ^{2}}{\left( 1-3\sin
^{2}\phi \right) ^{2}\left( 1-\sin ^{2}\phi \right) }\sin ^{2}\phi }{2\sigma
_{0}}\right) =0  \label{b5.31}
\end{eqnarray}%
where according to (\ref{a4.15}) the function $d\left( x\right) $ is
substituted by $\lambda _{0}^{2}f\left( x/\sigma _{0}\right) $.

Setting 
\begin{equation}
y=\sin ^{2}\phi  \label{b5.32}
\end{equation}%
and using designations (\ref{a4.24}), (\ref{a4.25}), we transform the
equation (\ref{b5.31}) to the form 
\begin{equation}
\varkappa \frac{a\left( 1-2y\right) ^{2}}{\left( 1-3y\right) ^{2}\left(
1-y\right) }+f\left( \nu \right) +f\left( \nu +2a\frac{\left( 1-2y\right)
\left( 3+y\right) }{\left( 1-3y\right) \left( 1-y\right) }\right) =0
\label{b5.33}
\end{equation}

The equations (\ref{a4.25a}) and (\ref{b5.33}) form a system of two
necessary conditions, imposed on parmeters of the helical world chain. Each
link of the chain consists of two vectors: leading vector $\mathbf{P}_{s}%
\mathbf{P}_{s+1}$ and stabilizing vector $\mathbf{P}_{s}\mathbf{Q}_{s}$.
Parameter $\varkappa =\sigma _{0}/\lambda _{0}^{2}$ is determined by the
space-time geometry. The quantity $\nu =-2l^{2}/\sigma _{0}$ describes the
length of the spacelike leading vector $\mathbf{P}_{s}\mathbf{P}_{s+1}$.
Parameter $a/y=2l_{1}^{2}/\sigma _{0}$ describes the length of the
projection of the leading vector $\mathbf{P}_{s}\mathbf{P}_{s+1}$ on the
plane of rotation. Finally, $y=\sin ^{2}\phi $ describes the angle $2\phi $
of rotation of the leading vector in the plane of rotation.

Numerical solutions of equations (\ref{a4.25a}) and (\ref{b5.33}) are
presented for the parameter $\varkappa =1$

\begin{equation*}
\begin{array}{llll}
\nu & a & y & s_{\perp }^{2}/\sigma _{0} \\ 
0.1 & 4.191\,5\times 10^{-2} & 0.392\,41 & 0.129\,57 \\ 
0.15 & 0.106\,61 & 0.444\,36 & 2.409\,8\times 10^{-2} \\ 
0.2 & 0.192\,36 & 0.462\,67 & 1.432\,4\times 10^{-2} \\ 
0.3 & 0.401\,37 & 0.474\,61 & 1.155\,3\times 10^{-2} \\ 
0.4 & 0.637\,01 & 0.478\,89 & 1.193\,1\times 10^{-2} \\ 
0.5 & 0.5 & 
\begin{array}{l}
0.468\,09 \\ 
0.398\,99%
\end{array}
& 
\begin{array}{l}
2.502\,3\times 10^{-2} \\ 
1.096\,7%
\end{array}
\\ 
0.6 & 0.136 & 0.406\,67 & 0.202\,86 \\ 
0.615 & 6.956\,7\times 10^{-2} & 0.375\,28 & 0.582\,94%
\end{array}%
\end{equation*}

\section{Estimation of wobbling of leading vector}

Solutions of equations, which describe the necessary conditions of the fact,
that the world chain may be a helix, are not unique. There may be solutions
of (\ref{a5.3a}), described by nonvanishing $\alpha $ and $\beta $, which
generate wobbling and violate the helical character of world chain. We write
six equation (\ref{a5.3a}) as equation for $\alpha ,\beta $ with parameters $%
l,q,s,\rho $, satisfying the necessary conditions (\ref{a4.25a}) and (\ref%
{b5.33}). We obtain instead of equations (\ref{b5.11}), (\ref{b5.12}) the
following two equations%
\begin{equation}
\alpha ^{2}+2\left( q.\alpha \right) =0  \label{b6.1}
\end{equation}%
\begin{equation}
\left( l.\alpha \right) \left( 1+\frac{\lambda _{0}^{2}}{\sigma _{0}}%
f^{\prime }\left( \frac{2l^{2}+2l_{1}^{2}\sin ^{2}\phi }{\sigma _{0}}\right)
\right) =0  \label{b6.2}
\end{equation}%
where the quantities $l,q$ satisfy the necessary conditions (\ref{b5.33}) (%
\ref{a4.25a}), and $f^{\prime }$ is a derivative of the function (\ref{a4.26}%
), which is always nonnegative. Then it follows from (\ref{b6.2}) 
\begin{equation}
\left( l.\alpha \right) =l_{0}\alpha _{0}-l_{1}\alpha _{1}=0  \label{b6.3}
\end{equation}

Equations (\ref{b6.1}), (\ref{b6.3}) contain only the variable $\alpha $
(but not $\beta $) and coincide with the equations (\ref{a4.41}), (\ref%
{a4.42}). However, there are additional constraints, containing $\alpha $.
As a result the constraints on $\alpha $ distinguish from the relation (\ref%
{a4.46}), describing values of $\alpha $ without the stabilizing vector $%
\mathbf{P}_{s}\mathbf{Q}_{s}$.

In the developed form the relations (\ref{b5.16}), (\ref{b5.17}) have the
form 
\begin{equation}
\beta _{0}^{2}-\mathbf{\beta }^{2}+2\left( s_{0}\beta _{0}-\beta
_{1}s_{\perp }\cos \phi \left( 1-4\sin ^{2}\phi \right) -\beta _{2}s_{\perp
}\sin \phi \left( 1+2\cos \left( 2\phi \right) \right) \right) =0
\label{b6.4}
\end{equation}%
\begin{equation}
\left( \beta _{0}s_{0}-\mathbf{\beta s}\right) \left( 1+\frac{\lambda
_{0}^{2}}{\sigma _{0}}f^{\prime }\left( \frac{\left( \rho +l\right) ^{2}}{%
2\sigma _{0}}\right) \right) -\frac{\lambda _{0}^{2}}{\sigma _{0}}f^{\prime
}\left( \frac{\left( \rho +l\right) ^{2}}{2\sigma _{0}}\right) \left(
l.\beta \right) =0  \label{b6.5}
\end{equation}%
They contain only the variable $\beta $ (but not $\alpha $)

Finally the relations (\ref{b5.23}), (\ref{b5.24}) in the developed form can
be written as follows%
\begin{eqnarray}
&&\gamma _{0}^{2}-\mathbf{\gamma }^{2}+2\left( s_{0}-l_{0}\right) \gamma
_{0}-\gamma _{1}\left( s_{\perp }\cos \phi \left( 1-4\sin ^{2}\phi \right)
-l_{1}\cos \left( 2\phi \right) \right)  \notag \\
&&-2\gamma _{2}\left( s_{\perp }\sin \phi \left( 1+2\cos \left( 2\phi
\right) \right) +l_{1}\sin \left( 2\phi \right) \right) =0  \label{b6.6}
\end{eqnarray}%
\begin{equation}
l_{0}\beta _{0}-l_{1}\beta _{1}+s_{0}\alpha _{0}-\mathbf{s\alpha }=0
\label{b6.7}
\end{equation}%
The relation (\ref{b6.7}) is a linear combination of equations (\ref{b5.17})
and (\ref{b5.24}), which does not contain the function $f$. Relations (\ref%
{b6.6}) and (\ref{b6.7}) contain both quantities $\alpha ,\beta $ and $%
\gamma =\beta -\alpha $. The constraints (\ref{b6.6}) and (\ref{b6.7})
modify the constraints (\ref{a4.46}), transforming the hyperboloid into
ellipsoid.

We suppose for simplicity, that the vector $\mathbf{P}_{s}\mathbf{Q}_{s}$ is
very long ($s_{0}\gg \sigma _{0}$). We suppose, that $s_{0}\rightarrow
\infty $. In this case we obtain from the relation (\ref{b6.5}), that $\beta
_{0}=0$. It follows from (\ref{b6.7}), that $\alpha _{0}=0$. Besides, it
follows from (\ref{b6.3}), that $\alpha _{1}=0$. Thus, solutions of the
equations (\ref{b6.5}), (\ref{b6.7}) and (\ref{b6.3}) have the form 
\begin{equation}
\beta _{0}=\alpha _{0}=0,\qquad \alpha _{1}=0  \label{b6.8}
\end{equation}

At the constraints (\ref{b6.8}) three other equations (\ref{b6.1}), (\ref%
{b6.4}) and (\ref{b6.6}) take the form 
\begin{equation}
\alpha _{2}^{2}+\alpha _{3}^{2}+2l_{1}\sin \left( 2\phi \right) \alpha _{2}=0
\label{b6.9}
\end{equation}%
\begin{equation}
\beta _{1}^{2}+\beta _{2}^{2}+\beta _{3}^{2}+2\beta _{1}s_{\perp }\cos \phi
\left( 1-4\sin ^{2}\phi \right) -2\beta _{2}s_{\perp }\sin \phi \left(
1+2\cos \left( 2\phi \right) \right) =0  \label{b6.10}
\end{equation}%
\begin{eqnarray}
&&\gamma _{1}^{2}+\gamma _{2}^{2}+\gamma _{3}^{2}+2\gamma _{1}\left(
s_{\perp }\cos \phi \left( 1-4\sin ^{2}\phi \right) -l_{1}\cos \left( 2\phi
\right) \right)  \notag \\
&&+2\gamma _{2}\left( s_{\perp }\sin \phi \left( 1+2\cos \left( 2\phi
\right) \right) -l_{1}\sin \left( 2\phi \right) \right) =0  \label{b6.11}
\end{eqnarray}%
Solution of equation (\ref{b6.9}) has the form 
\begin{equation}
\alpha _{1}=0,\qquad \alpha _{2}=-l_{1}\sin \left( 2\phi \right) \left(
1-\cos \eta \right) ,\qquad \alpha _{3}=l_{1}\sin \left( 2\phi \right) \sin
\eta  \label{b6.12}
\end{equation}%
where $\eta $ is an arbitrary angle.

Solution of equation (\ref{b6.10}) has the form%
\begin{eqnarray}
\beta _{1} &=&-s_{\perp }\cos \phi \left( 1-4\sin ^{2}\phi \right) +s_{\perp
}\cos \xi _{1}\cos \xi _{2}  \label{b6.14a} \\
\beta _{2} &=&-s_{\perp }\sin \phi \left( 1+2\cos \left( 2\phi \right)
\right) +s_{\perp }\cos \xi _{1}\sin \xi _{2}  \label{b6.14b} \\
\beta _{3} &=&s_{\perp }\sin \xi _{1}  \label{b6.14c}
\end{eqnarray}%
where the quantities $\xi _{1},$ $\xi _{2}$ are arbitrary. and the quantity $%
s_{\perp }$ is determined by the relation (\ref{b5.30}).

Substituting (\ref{b6.12}) - (\ref{b6.14c}) in (\ref{b6.11}), one obtains a
constraint on the quantities $\eta ,$ $\xi _{1},$ $\xi _{2}$. Independently
of this constraint the 3-vector 
\begin{equation}
\mathbf{q}+\mathbf{\alpha }=\left\{ l_{1}\cos \left( 2\phi \right)
,l_{1}\sin \left( 2\phi \right) \cos \eta ,l_{1}\sin \left( 2\phi \right)
\sin \eta \right\}  \label{c5.17}
\end{equation}%
has the same 3-length $l_{1}$, as the length of 3-vector $\mathbf{l}=\left\{
l_{1},0,0\right\} $. The angle between the 3-vectors $\mathbf{q}+\mathbf{%
\alpha }$ and $\mathbf{l}\ $is equal to $2\phi $. If $\eta =0$, then $\alpha
=0$, and vectors $\mathbf{P}_{s}\mathbf{P}_{s+1}$ and $\mathbf{P}_{s+1}%
\mathbf{P}_{s+2}$ are elements of the same helix.

We see that the stabilizing vector $\mathbf{P}_{s}\mathbf{Q}_{s}$ reduces
wobbling of vector $\mathbf{P}_{s}\mathbf{P}_{s+1}$. In the case of equation
(\ref{a4.46}) the spatial component $\mathbf{\alpha }$ of the 4-vector $%
\alpha $ may be infinite. In the case of the equation (\ref{b6.9}) the
length $\left\vert \mathbf{\alpha }\right\vert $ of the spatial component $%
\mathbf{\alpha }$ of the 4-vector $\alpha $ is less, than $\left\vert
l_{1}\sin \left( 2\phi \right) \right\vert .$ Thus, the stabilizing vector $%
\mathbf{P}_{s}\mathbf{Q}_{s}$ reduces the wobbling of the world chain. One
cannot be sure, that this wobbling does not destroy the helical character of
the world chain. However, The main question is, whether or not the evolution
of the world chain in the spacelike direction lead to the world chain, which
is timelike on the average.

Any next point $P_{l}$ of the world chain jumps along the timelike direction
at the distance $l_{0}$ and in the 3-space, which is orthogonal to this
direction, the point jumps at the distance $l_{1}>l_{0}$. Direction of the
jump in the 3-space is described by the vector $\mathbf{q}+\mathbf{\alpha }$%
, which is given by the relation (\ref{c5.17}). The length of $\mathbf{q}+%
\mathbf{\alpha }$ is $l_{1}$. If the direction of jump is completely random,
the displacement $L_{n}$ for $n$ steps ($n\gg 1)$ is proportional to $\sqrt{n%
}l_{1}$, whereas displacement in the temporal direction is $nl_{0}$. It
means that the mean velocity 
\begin{equation*}
\left\langle v\right\rangle =\frac{\sqrt{n}l_{1}}{nl_{0}}=\frac{l_{1}}{\sqrt{%
n}l_{0}}<1,\qquad n\gg 1
\end{equation*}%
tends to zero for $n\rightarrow \infty $, although $l_{0}<l_{1}$. In the
case, if $\alpha =0$ and the 3-vector $\mathbf{q}+\mathbf{\alpha }$\textbf{\ 
}determined by (\ref{c5.17}) is not random, the world chain form a helix
with timelike axis. In this case the mean velocity tends to zero also. It
should expect that in the case, when the vector (\ref{c5.17}) is stochastic,
but its stochasticity is restricted by the relation (\ref{c5.17}) (the angle 
$\eta $ is completely random), the mean world chain will be also timelike on
the average. We cannot prove this fact strictly now, but this result seems
to be very probable.

\section{Discussion}

The obtained classical helical world chain (\ref{a4.6}) associates with the
classical Dirac particle, which has alike world line (\ref{e6.33}), (\ref%
{e6.34}). The direction of the mean momentum distinguishes from the
direction of the 4-velocity. This fact is characteristic for both particles
(the Dirac particle, and the particle, described by the world chain (\ref%
{a4.6})). Both particles have angular moment. For the Dirac particle the
mass $m$, which enters in the Dirac equation, distinguishes from the mass $M$
of the particle moving along the world line (\ref{e6.33}), (\ref{e6.34}) 
\cite{R2004}. As to the mass of the particle, described by the world chain (%
\ref{a4.6}), it is not yet determined. For determination of the mass, one
needs to consider the world chain (\ref{a4.6}) of charged particle with the
skeleton $\left\{ P_{k},P_{k+1},Q_{k+1}\right\} $ in the distorted
space-time of Klein-Kaluza, containing electromagnetic field.

Existence of helical world chain with timelike axis seems to be rather
unexpected, because leading vectors $\mathbf{P}_{k}\mathbf{P}_{k+1}$ of the
chain are spacelike, and it corresponds to superluminal motion of a
particle. Superluminal motion seems to be incompatible with the relativity
principle, which admits only motion with the speed less, than the speed of
the light. However, this constraint is valid only for continuous space-time
geometry, which admits unlimited divisibility. In a discrete geometry there
are no distances less, than some elementary length, and it is difficult to
formulate the relativity principle statement on impossibility of
superluminal motion. One needs another more adequate formulation of the
relativity principle.

Is the space-time geometry (\ref{a4.14}) discrete? At $\sigma
_{0}\rightarrow 0$ the space-time geometry (\ref{a4.14}) turns to the
space-time geometry%
\begin{equation}
\sigma _{\mathrm{d}}=\sigma _{\mathrm{M}}+\lambda _{0}^{2}\mathrm{sgn}\left(
\sigma _{\mathrm{M}}\right)  \label{a7.1}
\end{equation}%
which is certainly discrete, because in the space-time there no timelike
intervals $\left\vert \mathbf{PQ}\right\vert $, which are less, than $%
\lambda _{0}$, and there are no spacelike intervals $\sqrt{-\left\vert 
\mathbf{PQ}\right\vert ^{2}}$, which are less, than $\lambda _{0}$. In such
a space-time geometry there are no particles, whose geometrical mass $\mu $
is less than $\lambda _{0}$.

However, if $\sigma _{0}>0$, is the space-time geometry discrete? To answer
this question, we introduce the parameter of discreteness: the relative
density of points in the space-time with respect to the point density in the
space-time of Minkowski. Let us define the quantity $\rho \left( \sigma _{%
\mathrm{d}}\right) $ by means of the relation 
\begin{equation}
\rho \left( \sigma _{\mathrm{d}}\right) =\frac{d\sigma _{\mathrm{M}}\left(
\sigma _{\mathrm{d}}\right) }{d\sigma _{\mathrm{d}}}  \label{a7.2}
\end{equation}%
In the case (\ref{a4.14}) we have for $\sigma _{\mathrm{M}}\in \left[
-\sigma _{0},\sigma _{0}\right] $ 
\begin{equation}
\sigma _{\mathrm{M}}+\lambda _{0}^{2}\left( \frac{\sigma _{\mathrm{M}}}{%
\sigma _{0}}\right) ^{3}-\sigma _{\mathrm{d}}=0  \label{a7.3}
\end{equation}%
Resolving (\ref{a7.3}) with respect to $\sigma _{\mathrm{M}}$, we obtain 
\begin{equation}
\sigma _{\mathrm{M}}=\sigma _{0}g^{1/3}\left( \sigma _{\mathrm{d}}\right) -%
\frac{\sigma _{0}^{2}}{3\lambda _{0}^{2}}g^{-1/3}\left( \sigma _{\mathrm{d}%
}\right)  \label{a7.4}
\end{equation}%
where%
\begin{equation}
g\left( \sigma _{\mathrm{d}}\right) =\sqrt{\frac{\sigma _{0}^{3}}{27\lambda
_{0}^{6}}+\frac{\sigma _{d}^{2}}{4\lambda _{0}^{4}}}+\frac{\sigma _{d}}{%
2\lambda _{0}^{2}}  \label{a7.5}
\end{equation}

Taking into account (\ref{a7.4}) we obtain the world function $\sigma _{%
\mathrm{M}}$ as a function of $\sigma _{\mathrm{d}}$%
\begin{equation}
\sigma _{\mathrm{M}}=\left\{ 
\begin{array}{lll}
\sigma _{\mathrm{d}}-\lambda _{0}^{2}\mathrm{sgn}\left( \sigma _{\mathrm{d}%
}\right) & \text{if} & \left\vert \sigma _{\mathrm{d}}\right\vert >\sigma
_{0}+\lambda _{0}^{2} \\ 
\sigma _{0}g^{1/3}\left( \sigma _{\mathrm{d}}\right) -\frac{\sigma _{0}^{2}}{%
3\lambda _{0}^{2}}g^{-1/3}\left( \sigma _{\mathrm{d}}\right) & \text{if} & 
\left\vert \sigma _{\mathrm{d}}\right\vert \leq \sigma _{0}+\lambda _{0}^{2}%
\end{array}%
\right.  \label{a7.6}
\end{equation}%
The relative density of points in the space-time geometry $\mathcal{G}_{%
\mathrm{d}}$ with respect to the standard geometry $\mathcal{G}_{\mathrm{M}}$
of Minkowski is given by the relation (\ref{a7.2}). The expression for $\rho
\left( \sigma _{\mathrm{d}}\right) $ is given by the relation%
\begin{equation}
\rho \left( \sigma _{\mathrm{d}}\right) =\left\{ 
\begin{array}{lll}
1 & \text{if} & \left\vert \sigma _{\mathrm{d}}\right\vert >\sigma
_{0}+\lambda _{0}^{2} \\ 
g^{\prime }\left( \sigma _{\mathrm{d}}\right) \left( \frac{\sigma _{0}}{3}%
g^{-2/3}\left( \sigma _{\mathrm{d}}\right) +\frac{\sigma _{0}^{2}}{9\lambda
_{0}^{2}}g^{-4/3}\left( \sigma _{\mathrm{d}}\right) \right) & \text{if} & 
\left\vert \sigma _{\mathrm{d}}\right\vert \leq \sigma _{0}+\lambda _{0}^{2}%
\end{array}%
\right.  \label{a7.7}
\end{equation}%
where $g^{\prime }\left( \sigma _{\mathrm{d}}\right) $ is given by the
relation 
\begin{equation}
g^{\prime }\left( \sigma _{\mathrm{d}}\right) =\frac{\sigma _{d}}{4\lambda
_{0}^{4}\sqrt{\frac{1}{27}\frac{\sigma _{0}^{3}}{\lambda _{0}^{6}}+\frac{1}{%
4\lambda _{0}^{4}}\sigma _{d}^{2}}}+\frac{1}{2\lambda _{0}^{2}}  \label{a7.8}
\end{equation}%
If $\sigma _{0}\rightarrow 0$ and $\sigma _{0}\ll \lambda _{0}^{2}$, we have
approximately%
\begin{equation}
g\left( \sigma _{\mathrm{d}}\right) =\frac{\sigma _{d}}{\lambda _{0}^{2}},
\qquad g^{\prime }\left( \sigma _{\mathrm{d}}\right) =\frac{1}{\lambda
_{0}^{2}}  \label{a7.9}
\end{equation}%
In the limit $\sigma _{0}\rightarrow 0$, when the world function (\ref{a4.14}%
) turns into the world function (\ref{a7.1}) of the completely discrete
geometry, we obtain for the relative density%
\begin{equation}
\lim_{\sigma _{0}\rightarrow 0}\rho \left( \sigma _{\mathrm{d}}\right)
=\left\{ 
\begin{array}{lll}
1 & \text{if} & \left\vert \sigma _{\mathrm{d}}\right\vert >\lambda _{0}^{2}
\\ 
0 & \text{if} & \left\vert \sigma _{\mathrm{d}}\right\vert \leq \lambda
_{0}^{2}%
\end{array}%
\right.  \label{a7.10}
\end{equation}

Thus, $\rho \left( \sigma _{\mathrm{d}}\right) =0$ for $\sigma _{\mathrm{d}%
}\in \left( -\lambda _{0}^{2},\lambda _{0}^{2}\right) $, and this fact
correspond to the space-time geometry (\ref{a7.1}), where close points, for
which $\left\vert \sigma _{\mathrm{d}}\right\vert \leq \lambda _{0}^{2}$,
are absent. The relative density $\rho \left( \sigma _{\mathrm{d}}\right) $
of points may serve as quantity, describing the discreteness of the
space-time geometry and the character of this discreteness. The discreteness
may be complete, when the density $\rho \left( \sigma _{\mathrm{d}}\right) \ 
$vanishes in some region as in the case (\ref{a7.10}). But the discreteness
may be incomplete, as in the case (\ref{a7.7}). In this case for $\sigma
_{0}=\lambda _{0}^{2}$ we have 
\begin{equation}
\rho \left( \sigma _{\mathrm{d}}\right) =\frac{1}{6\sqrt{\frac{4}{27}\lambda
_{0}^{4}+\sigma _{d}^{2}}}\left( \sqrt[3]{g_{1}\left( \sigma _{\mathrm{d}%
}\right) }+\frac{1}{3}\frac{1}{\sqrt[3]{g_{1}\left( \sigma _{\mathrm{d}%
}\right) }}\right) ,\qquad \sigma _{\mathrm{d}}\in \left( -2\lambda
_{0}^{2},2\lambda _{0}^{2}\right)  \label{a7.11}
\end{equation}%
where%
\begin{equation}
g_{1}\left( \sigma _{\mathrm{d}}\right) =\frac{1}{2\lambda _{0}^{2}}\left(
\sigma _{d}+\sqrt{\frac{4}{27}\lambda _{0}^{4}+\sigma _{d}^{2}}\right)
\label{a7.12}
\end{equation}%
The expression (\ref{a7.11}) is a symmetric function of $\sigma _{d}$, as
one can see from (\ref{a7.3}). It is symmetric, indeed, although it does not
look formally as a symmetric function of $\sigma _{\mathrm{d}}$. Numerical
values of $\rho \left( \sigma _{\mathrm{d}}\right) ,$ $\sigma _{\mathrm{d}%
}\in \left( -2\lambda _{0}^{2},2\lambda _{0}^{2}\right) $ are presented in
the table 
\begin{equation*}
\begin{array}{ll}
\sigma _{d} & \rho \\ 
0 & 0.5 \\ 
0.2\lambda _{0}^{2} & 0.449\,82 \\ 
0.4\lambda _{0}^{2} & 0.362\,72 \\ 
0.8\lambda _{0}^{2} & 0.243\,63%
\end{array}%
\qquad 
\begin{array}{ll}
\sigma _{d} & \rho \\ 
1.0\lambda _{0}^{2} & 0.208\,62 \\ 
1.2\lambda _{0}^{2} & 0.182\,85 \\ 
1.4\lambda _{0}^{2} & 0.163\,20 \\ 
1.6\lambda _{0}^{2} & 0.147\,74%
\end{array}%
\qquad 
\begin{array}{ll}
\sigma _{d} & \rho \\ 
1.8\lambda _{0}^{2} & 0.135\,28 \\ 
2.0\lambda _{0}^{2} & 0.125 \\ 
-0.2\lambda _{0}^{2} & 0.449\,82 \\ 
-0.4\lambda _{0}^{2} & 0.362\,72%
\end{array}%
\end{equation*}%
The relative density $\rho \left( \sigma _{\mathrm{d}}\right) $ is less,
than unity. It may be interpreted in the sense, that the space-time geometry
is discrete only partly. Nevertheless the incompletely discrete space-time
geometry discriminates most of world chains with spatial leading vector,
remaining only some of them.

Multivariance of particle motion and discrimination of some states of motion
play the crucial role in structure of elementary particles, as well as in
the structure of atoms. Let us explain this circumstance in the example of
the hydrogen atom. According to laws of the classical mechanics the electron
of the hydrogen atom is to fall onto the nucleus due to the Coulomb
attraction. Two reasons prevent from this falling: (1) multivariant
(stochastic) motion of the electron, and (1) rotation of electron around the
nucleus.

The multivariant motion of the electron leads to escape of the electron from
the nuclear surface. This process has the same nature, as an escape of dust
from the Earth surface. Moving multivariantly (as Brownian particles), the
flecks of dust form a stationary distribution in the gravitational field of
the Earth. If multivariance of their motion is cut out, the flecks of the
dust fall onto the surface of the Earth. Statistical description of the
electron distribution and the dust distribution are different, because the
multivariant electron motion is conceptually relativistic, whereas the
Brownian particles motion is nonrelativistic. One may describe Brownian
particles by means of probabilistic statistical description, whereas one may
use only dynamical conception of statistical description for statistical
description of multivariant motion of relativistic particles.

Rotation of the electron around the nucleus creates the field of centrifugal
force, which is added to the Coulomb force. As a result additional
distributions of the electrons appear. If the obtained distribution of
electrons is nonstationary, the electrons emanate the electromagnetic
radiation until the electron distribution becomes to be stationary. Thus,
the electromagnetic radiation carries out discrimination of nonstationary
states (electron distributions). The multivariance of the electron motion
and mechanism of discrimination of non-stationary states generates the
structure of the hydrogen atom and discrete character of the radiation
spectra. From the mathematical viewpoint the discrete character of the
electron states is conditioned by procedure of the eigenstates
determination. Only eigenstates of the Hamilton operator appear to be
stationary and stable.

The multivariance of the particle motion and some mechanism of
discrimination play also the crucial role in the understanding of the
structure of elementary particles. However, in the case of the elementary
particle structure the discrimination mechanism is conditioned by some
metric (geometric) forces, which appear, when we use space-time geometry of
Minkowski instead of the real multivariant space-time geometry. Formally
these forces have the form of additional terms of the type (\ref{a4.18}) in
dynamic equations. These additional terms are expressed via the space-time
distortion $d$. They describe both multivariance of motion and the
discrimination mechanism. The multivariance of motion is associated with the
multivariance of the vector equivalence definition (\ref{a1.3}), whereas the
discrimination mechanism is associated with the zero-variance of the same
definition (\ref{a1.3}) for some vectors. Besides, as we have seen, the
zero-variance (discrimination) is associated with the discreteness (or
partial discreteness) of the space-time geometry.

It is very important, that consideration of multivariant space-time geometry 
\textit{is not a hypothesis}, which needs an experimental test.
Consideration of the multivariant space-time geometry is a corollary of
correction of our imperfect conception of geometry. Conception of geometry,
based on supposition that any space-time geometry may be axiomatized (i.e.
may be concluded from some system of axioms), is imperfect, because it does
not admit one to construct multivariant geometry conceptually. However, the
motion of electrons and other elementary particles is multivariant.
Multivariance of this motion is an experimental fact, which cannot be
ignored. As far as the imperfect conception of geometry did not admit one to
construct multivariant space-time geometry, investigators were forced to
ascribe multivariance to dynamics, introducing quantum principles with all
their attributes.

The quantum principles look enigmatic and artificial, because multivariance
is ascribed to dynamics, whereas it should be ascribed to the space-time
geometry. Multivariance and zero-variance as properties of the space-time
geometry look as quite natural properties of the definition (\ref{a1.3}).
Indeed, it does not follow from anywhere, that equations (\ref{a1.3}) are to
have unique solution for arbitrary world function, which determines the form
of these equations. Absence of any hypotheses is a very important property
of the geometrical approach to the structure of elementary particles.
Besides, the geometrical dynamics is very general and simple. Dynamic
equations of the geometric dynamics do not use even differential equations.
Formulation of dynamic equations does not contain a reference to the
coordinate system. On the other hand, when the geometric dynamics in the
real space-time is described in terms of the space-time of Minkowski, one
obtains additional metric forces, which look rather exotic. They can be
obtained hardly in the framework of the conventional approach.

The conventional approach to the theory of elementary particles contains a
lot of secondary concepts and properties. One may not see any discrimination
mechanism in wave functions, field equations, branes, symmetries and other
remote corollaries of the unknown structure of elementary particles. But it
is impossible to obtain and to understand the discrete properties of
elementary particles without a reliable mechanism of discrimination.

Even if investigating and systematizing these remote corollaries, one
succeeds to obtain a perfect systematization of elementary particles, one
can obtain structure of elementary particles from the perfect
systematization with the same success, as one can obtain the atomic
structure from the periodical system of chemical elements.

\section{Concluding remarks}

Consideration of T-geometry as a space-time geometry admits one to obtain
dynamics of a particle as corollary of its geometrical structure. Evolution
of the geometrical object in the space-time is determined by the skeleton $%
\left\{ P_{0},P_{1},..P_{n}\right\} $ of the geometrical object and by
fixing of the leading vector $\mathbf{P}_{0}\mathbf{P}_{1}$. The skeleton
and the leading vector determine the world chain, which describes the
evolution completely. The world chain may wobble, it is manifestation of the
space-time geometry multivariance. Quantum effects are only one of
manifestation of the multivariance. It is remarkable, that for determination
of the world chain one does not need differential equations, which may be
used only on the space-time manifold. One does not need space-time
continuity (continual geometry). Of course, one may introduce the continual
coordinate system and write dynamic differential equation there. One may,
but it is not necessary. In general, the geometrical dynamics (i.e. dynamics
generated by the space-time geometry) is a discrete dynamics, where step of
evolution is determined by the length of the leading vector. It is possible,
that one will need a development of special mathematical technique for the
geometrical dynamics.

The real space-time geometry contains the quantum constant $\hbar $ as a
parameter. As a result the geometric dynamics explains freely quantum
effects, but not only them. The particle mass is geometrized (the particle
mass is simply a length of some vector). As a result the problem of mass of
elementary particles is simply a geometrical problem. It is a problem of the
structure of elementary geometrical object and its evolution. One needs
simply to investigate different forms of skeletons of simplest geometrical
objects. In general, not all skeletons are possible, because at the spatial
evolution the world chain is observable (helical) only for several
skeletons. Additional points of skeleton lead to additional (sometimes
unexpected) properties of corresponding elementary geometrical objects
(elementary particles).

Note that the geometric dynamics does not contain a rotational motion. It
contains only a shift. All vectors of the skeleton $\left\{ P_{0}^{\left(
s\right) },P_{1}^{\left( s\right) },..P_{n}^{\left( s\right) }\right\} $ of
the link $L_{s}$ are equivalent to vectors of the skeleton $\left\{
P_{0}^{\left( s+1\right) },P_{1}^{\left( s+1\right) },...\right. \left.
P_{n}^{\left( s+1\right) }\right\} $ of the adjacent link $L_{s+1}$. Such a
situation is quite reasonable, because the geometrical dynamics describes
evolution of free particles. The rotating particle cannot be completely
free, because in the rotating particle there is centripetal acceleration.
However, acceleration of all parts of the body has to be absent for
completely free motion. On the other hand, the geometric dynamics contains
the spatial evolution, which absent in the conventional dynamics. From the
geometrical viewpoint we may not discriminate spatial evolution on the
basis, that the leading vector $\mathbf{P}_{0}\mathbf{P}_{1}$ is spacelike
and its length is imaginary. In fact the spatial evolution discriminates
itself, by the fact, that the corresponding world chain is unobservable, in
general. It appears to be observable only for some complex skeletons,
consisting of more, than two points. The world chain, describing the spatial
evolution is observable only in the case, when it may be localized near the
world chain of the observer. It takes place, when the world chain has a
shape of a helix with timelike axis, or some other shape, which may be
localized near the world chain of the observer. As a result not all
skeletons appear to be observable.

Although the geometric dynamics does not contain a rotation, but the
corollaries of the rotation (angular momentum, magnetic momentum) may be
obtained as a result of the spatial evolution, when the world chain is a
helix. Apparently, the fact, that such a "particle rotation" is a corollary
of the spatial evolution, leads to the spin discreteness of the Dirac
particle. Of course, such statements are to be tested by exact mathematical
investigations of different types of skeletons and of different space-time
geometries. However, such a statement of the problem is quite concrete and
realizable.

Note, that the geometric dynamics in the real (non-Minkowskian) space-time
contains additional terms with respect to dynamics in the space-time of
Minkowski. From viewpoint of the space-time of Minkowski these additional
terms may be interpreted as some (metric) interactions, which take place
inside the elementary particles. From the conventional viewpoint these
interactions look very exotic and strange. It is impossible (or very
difficult) to guess at them, starting from conventional conception of the
space-time and dynamics. In the geometric dynamics there are no additional
interactions, if we use the true space-time geometry. However, additional
interactions appear, if we use inadequate geometry (for instance, geometry
of Minkowski, or Riemannian geometry). In other words, it is possible to
compensate false space-time geometry by introduction of additional
interactions. It is well known from the general relativity, that the motion
of free body in the curved space-time looks as a motion in the gravitational
field, if one interprets this motion as a motion in the space-time of
Minkowski.

Description of conceptually new unknown phenomena by means of a change of
the space-time geometry is simpler, than an introduction of additional
interactions, because the space-time geometry is described by the world
function, which is a function of two points. The form of the world function
for large distances is determined by the necessity of obtaining the
nonrelativistic quantum mechanics. Restrictions, imposed on the world
function at small distances, are determined by the condition, that the
spatial evolution may describe the Dirac particle. (Very many elementary
particles are the Dirac particles). Besides, the condition of localization
of the world chain (helical world chain) imposes restrictions on parameters
of the particle. Not all parameters of particles appear to be possible. This
condition is a condition of "peculiar quantization" of the particle
parameters, which include the particle mass.

Let us note that the contemporary theory of elementary particles returns to
geometrical considerations (strings, branes). However, these considerations
are restricted by the framework of the Riemannian geometries and geometries
close to the Riemannian geometry. For instance, the quantum geometry, which
uses operators instead of the point coordinates. This is some way of
introduction of multivariance in the geometry. However, this geometry is
developed on the basis of the linear vector space, which is a restriction on
the space-time geometry. In any case the conventional approach to the
space-time geometry considers only a part of all possible space-time
geometries. One cannot be sure, that the class of considered geometries
contains true space-time geometry. Of course, if one uses a false space-time
geometry, there is a possibility to correct the false space-time geometry by
means of additional interaction, generated by difference with the true
space-time geometry. But such a correction is difficult, especially if the
true geometry is discrete or close to the discrete geometry.

Note, that the geometry (\ref{a4.0}) is discrete, although it is given on
the continuous manifold of Minkowski. It is discrete, because the module of
distance between any two points is more, than $\lambda _{0}$. It is very
unexpected, because it is a common practice to consider any geometry on the
manifold as a continuous geometry, although in reality the geometry is
determined by the world function and only by the world function. A discrete
geometry is associated with a grid. Of course, a geometry, given on a grid,
cannot be continuous. However, a geometry, given on the continuous set of
points (manifold), may be discrete.

Why the microcosm physics of the twentieth century did leave the successful
program of the physics geometrization and choose the alternative program of
quantum theory? Discovery of the electron diffraction need of multivariance
of the microcosm physics. Multivariance may be taken into account either on
the level of the space-time geometry, or on the level of dynamics. The
multivariant space-time geometry was not known in the thirtieth, when the
electron diffraction was discovered. The nonrelativistic quantum mechanics
had been constructed already, and it was applied successfully for
explanation of the electron diffraction.

The space-time geometry is a basis of dynamics. Introducing multivariance in
dynamics, one can describe not only nonrelativistic phenomena of microcosm.
One can describe also relativistic phenomena and that part of the microcosm
physics, which is known as the theory of elementary particles. The
principles of quantum mechanics, which introduce multivariance in the
microcosm physics, were invented for the Newtonian conception of the
space-time, and their extrapolation to the relativistic phenomena appeared
to be problematic. Of course, some properties of the true space-time
geometry may be taken into account by introduction of additional
interactions. However, it is very difficult to invent and introduce
additional interactions without understanding of these innovations.
Capacities of the geometrical approach are very large, especially if one
takes into account all possible space-time geometries. The theory of
elementary particles returns to the geometrical description, but this
description is burthened by such concepts as wave function, string, brane,
which have very abstracted relation to the structure of elementary particles
and microcosm physics.


\begin{thebibliography}{99}
\bibitem{R2007a} Yu. A. Rylov, Deformation principle and further
geometrization of physics http://arXiv.org/abs/0704.3003

\bibitem{R90} Yu.A. Rylov, Extremal properties of Synge's world function and
discrete geometry. \textit{Journ. Math. Phys}. \textbf{31}, 2876-2890,
(1990).

\bibitem{R2001} Yu.A. Rylov, Geometry without topology as a new conception
of geometry. \textit{Int. Jour. Mat. \& Mat. Sci}. \textbf{30}, iss. 12,
733-760, (2002), (Available at http://arXiv.org/abs/math.MG/0103002 ).

\bibitem{R2005} Yu. A. Rylov, Tubular geometry construction as a reason for
new revision of the space-time conception. in \textit{What is Geometry?}
polimetrica Publisher, Italy, pp.201-235
http://www.polimetrica.com/polimetrica/406/

\bibitem{R2007b} Yu. A. Rylov, Different conceptions of Euclidean geometry. 
\textit{http://arXiv.org /abs /0709.2755}

\bibitem{R2004} Yu. A. Rylov, Is the Dirac particle composite? \textit{%
http://arXiv.org/abs/physics /0410045}

\bibitem{R2004b} Yu. A. Rylov, Is the Dirac particle completely
relativistic? \textit{http://arXiv.org /abs/physics /0412032}

\bibitem{R2001c} Yu.A. Rylov, Dynamic disquantization of Dirac equation. 
\textit{http://arXiv.org /abs/quant-ph/0104060}

\bibitem{R2005a} Yu. A. Rylov, Dynamical methods of investigation in
application to the Dirac particle. \textit{%
http://arXiv.org/abs/physics/0507084}

\bibitem{R2006} Yu. A. Rylov, Author's comments to referee's reports on the
paper by Y. A. Rylov "Dynamical methods of investigation in application to
the Dirac particle", submitted to a scientific journal. (Available at 
\textit{http://rsfq1.physics.sunysb. edu/\symbol{126}rylov /comm1e. pdf})

\bibitem{R91} Yu.A. Rylov, Non-Riemannian model of the space-time
responsible for quantum effects. \textit{Journ. Math. Phys}. \textbf{32(8)},
2092-2098, (1991).
\end{thebibliography}
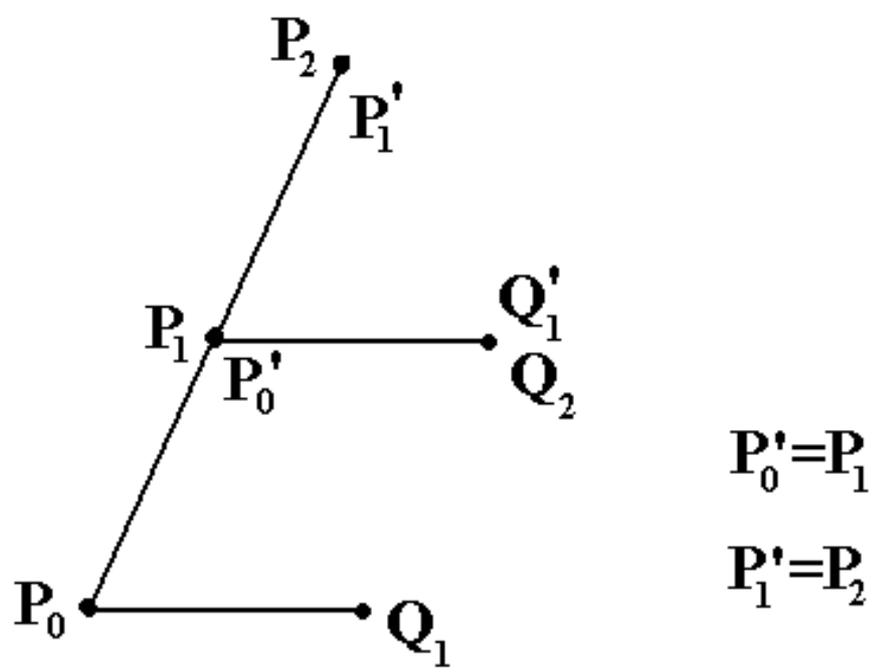
\end{document}